\documentclass[journal]{IEEEtran}

\usepackage{cite}
\usepackage{graphicx,subfigure,epsfig,graphics}
\usepackage{amsmath,amsthm,amsfonts,amssymb,mathrsfs}
\usepackage{multirow,dsfont,mathtools,bm}
\usepackage[dvipsnames]{xcolor}
\usepackage{colortbl}
\usepackage{tikz}
\usetikzlibrary{fit}
\usepackage{cleveref}
\usepackage{url}
\usepackage{breakurl}
\usepackage{soul}

\DeclarePairedDelimiter{\ceil}{\lceil}{\rceil}

\newcommand\addvmargin[1]{
  \node[fit=(current bounding box),inner ysep=#1,inner xsep=0]{};
}

\begin{document}
\title{Architecture and Algorithms for Privacy Preserving Thermal Inertial Load Management by \\A Load Serving Entity}
%
%
%

\author{Abhishek~Halder,~\IEEEmembership{Member,~IEEE,}
        Xinbo~Geng,~\IEEEmembership{Student Member,~IEEE,}
		 P.R.~Kumar,~\IEEEmembership{Fellow,~IEEE,}
        and~Le~Xie,~\IEEEmembership{Senior Member,~IEEE}
\thanks{The authors are with the Department
of Electrical and Computer Engineering, Texas A\&M University, College station,
TX, 77843 USA, e-mail: \texttt{\{ahalder,gengxbtamu,prk,le.xie\}@tamu.edu}. This material is based upon work partially supported by NSF under Contract Nos. ECCS-1546682, CPS-1239116, Science \& Technology Center Grant CCF-0939370, DGE-1303378, and ECCS-1150944.}
}

%
%

\maketitle

\begin{abstract}
Motivated by the growing importance of demand response in modern power system's operations, we propose an architecture and supporting algorithms for privacy preserving thermal inertial load management as a service provided by the load serving entity (LSE). We focus on an LSE managing a population of its customers' air conditioners, and propose a contractual model where the LSE guarantees quality of service to each customer in terms of keeping their indoor temperature trajectories within respective bands around the desired individual comfort temperatures. We show how the LSE can price the contracts  differentiated by the flexibility embodied by the width of the specified bands. We address architectural questions of (i) how the LSE can strategize its energy procurement based on price and ambient temperature forecasts, (ii) how an LSE can close the real time control loop at the aggregate level while providing individual comfort guarantees to loads, without ever measuring the states of an air conditioner for privacy reasons. Control algorithms to enable our proposed architecture are given, and their efficacy is demonstrated on real data. 
\end{abstract}


\IEEEpeerreviewmaketitle


\section{Introduction}

\IEEEPARstart{I}{n} traditional power systems, the role of feedback control has primarily been concentrated on the generation side where the demand side variability is dynamically matched via operating reserve. As more renewables are being integrated into the power grid to reduce carbon footprint, a new operational paradigm is emerging. Unlike fossil fuels, generation from renewables such as solar and wind, cannot be directly controlled. Thus, large scale renewable integration requires moving from control of the supply to control of the demand. This necessitates a collection of mechanisms, termed as demand response, for manipulating aggregate demand to offset the supply side variability. Thermal inertial loads such as air conditioners (ACs) are one of the primary candidates \cite{Callaway2011IEEE} for control on the demand side. They contribute to 48\% of U.S. annual residential consumption \cite{RECS2009}, and are potential enablers for demand response due to their ability to store energy and defer consumption without perceptible change in end-user functionality. 

In this paper, we consider an aggregator or load serving entity (LSE) managing a finite population of residential ACs through direct load control \cite{Callaway2011IEEE, CaramanisBohnSchweppe1982}. We address three distinct problems that are important for designing a direct load control architecture, viz. (1) \emph{how} to control a population of thermal inertial loads to guarantee quality of service such as comfort bounds for homes, (2) \emph{how} to achieve such control while preserving the privacy of the loads' states from the LSE, and (3) \emph{what} to control the aggregate consumption to, so that the cost of serving the collection of loads is minimized. Our strategy effectively produces demand response from the collection of homes according to how their temperatures are distributed within their comfort zones at any given time, without attempting to squeeze it from each and every home.

Holistically maintaining individual home privacy while still generating demand response that respects home comfort constraints does not seem to have been addressed in the literature. Previous works on thermal inertial load management \cite{ChongDebs1979,MalhameChong1985,Callaway2009Energy,BashashFathy2011ACC,MathieuKochCallaway2013TPS,Zhang2013TPS,GhaffariMouraKrstic2014DSMC,Meyn2015TAC} have focused on obtaining an aggregate model for a population of loads, and then used the resulting model, or an approximation of it, to track a given reference power trajectory.
The purpose of this paper is to present a comprehensive architecture and algorithms (Section \ref{ArchitectureStrategySection})  so that the LSE can manage a population of thermal inertial loads 
while addressing the three questions mentioned in the preceding paragraph. We address the first two problems by proposing a model-free setpoint velocity control algorithm (Section \ref{LoadControlSubsubsection}). We formulate and solve (Section \ref{OperationalPlanningSubsection}) the third as an open-loop optimal control problem. 
We also quantify (Section \ref{PerformanceSubsection}) the statistical performance of the LSE's strategy  under random perturbations that entail different real-time energy usage than planned in the day-ahead market. We show (Section \ref{ContractSectionLabel}) how the LSE can price the contracts differentiated by the flexibility provided by the customers with respect to their comfort requirements, and illustrate (Section \ref{CaseStudiesSectionLabel}) all our algorithms on day-ahead price forecast data from the Electric Reliability Council of Texas (ERCOT), and ambient temperature data from a weather station in Houston, Texas. Some preliminary results using these ideas applied to ERCOT real-time energy market price forecast data, appeared in \cite{AbhishekSmartgridComm2015}.

\subsubsection*{Notations}
The notation $\mathbf{x} \sim \rho$ means that the random vector $\mathbf{x}$ follows joint density function $\rho(x)$. We use the shorthand Lap($\kappa$) to denote the zero mean Laplace density function with scale parameter $\kappa > 0$, $\text{Gamma}(a,b)$ to denote the Gamma density with parameters $a, b>0$, $\text{Exp}(\lambda)$ to denote exponential density with rate $\lambda >0$, and $\mathcal{N}(\mu,\Sigma)$ to denote the Gaussian density with mean $\mu$ and covariance $\Sigma$. We use $\mathcal{N}\left(\mu,\sigma,a,b\right)$ to denote a truncated univariate Gaussian density with mean $\mu$ and standard deviation $\sigma$, supported on $[a,b]$. The symbols $\vee$ and $\wedge$ denote the maximum and minimum, respectively, and $[r]^{+} := 0 \vee r$, for any $r\in\mathbb{R}$. The notation $\mathds{1}_{S}$ denotes the indicator function of a set $S$, while the abbreviation i.i.d. stands for ``independent and identically distributed".


\section{Model}

\begin{figure}[hptb]
  \centering
    \includegraphics[width=0.49\textwidth]{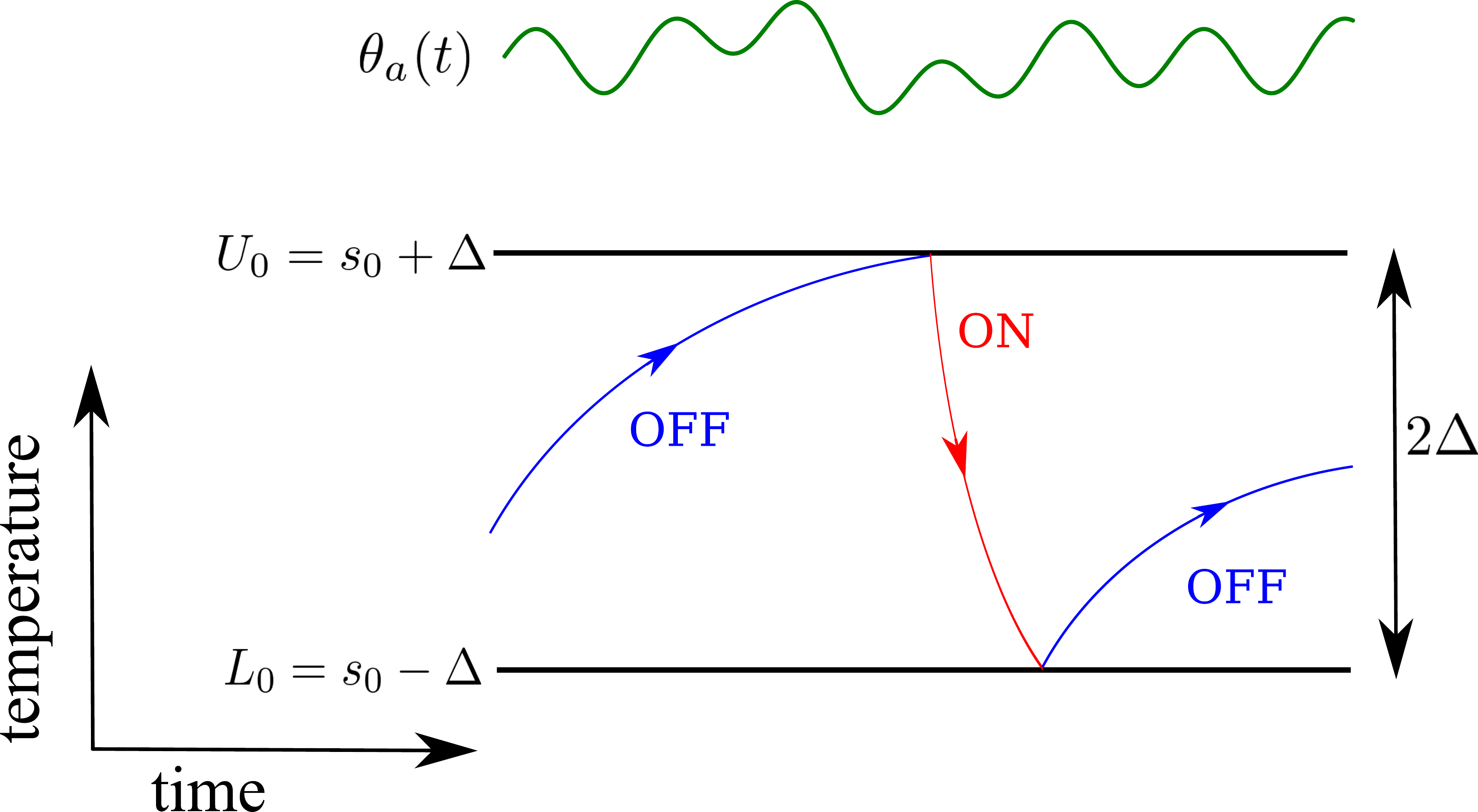}
    \caption{Thermal inertial dynamics of an AC, \emph{not} under direct load control, is shown with fixed setpoint $s_{0}$, and a comfort temperature interval $[L_{0},U_{0}]$ with range $2\Delta$, where $U_{0}=s_{0}+\Delta$ and $L_{0}=s_{0}-\Delta$. The ambient temperature trajectory (\emph{green}) is $\theta_{a}(t)$. The indoor temperature trajectory $\theta(t)\in[L_{0},U_{0}]$ consists of alternating OFF (\emph{blue, up-going}) and ON (\emph{red, down-going}) segments, as shown, where the boundaries $L_{0}$ and $U_{0}$ act as reflecting barriers for the ON and OFF segments, respectively.}
    \label{AirConditionerDynamics}
    \vspace*{-0.25in}
\end{figure}

\subsection{Dynamics of Thermal Inertial Load}
In this paper, we consider the ambient temperature is high enough that ACs are cooling homes, a situation of the sort where demand response is appropriate. 
We denote the indoor temperature at time $t$ by $\theta(t)$, and the ambient temperature by $\theta_{a}(t)$. At time $t=0$, an occupant privately sets a temperature $s_{0}$, called setpoint, with a willingness to tolerate at most $\pm \Delta$ temperature deviation from $s_{0}$, thus defining a temperature comfort range $[L_{0},U_{0}] := [s_{0}-\Delta,s_{0}+\Delta]$. The flexibility of an AC is measured by $2 \Delta$; larger the flexibility, greater the potential for demand response. The goal of the proposed architecture and algorithms is to enable the LSE exploit this flexibility without the LSE being aware of the load temperature $\theta(t)$ or the setpoint $s_0$, or even $\Delta$.

The dynamics of an AC \emph{not} under direct load control, is shown in Fig. \ref{AirConditionerDynamics}. When the AC is OFF, the indoor temperature $\theta(t)$ rises exponentially toward $\theta_{a}(t)$ until it hits the upper comfort limit $U_{0}$, at which point the AC turns ON. Once ON, $\theta(t)$ decreases exponentially until it hits the lower comfort limit $L_{0}$, at which point the AC turns OFF. 
While this qualitative behavior shown in Fig. \ref{AirConditionerDynamics} is valid for any ambient temperature trajectory satisfying $\theta_{a}(t) > U_{0}$, quantitatively speaking, temporal variations in $\theta_{a}(t)$ produce time-varying rates of heating (when AC is OFF) and cooling (when AC is ON). Let us denote the mode of an AC at time $t$ by $\sigma(t) = 1(0)$, signifying if it is ON (OFF). Its hysteretic switching trajectory is
\begin{eqnarray}
\sigma(t) := \begin{cases}
	1 \qquad\qquad\text{if}\quad\theta(t) = U_{0},\\
	0 \qquad\qquad\text{if}\quad\theta(t) = L_{0},\\
	\sigma\left(t^{-}\right)\quad\;\;\,\text{otherwise}. 
 \end{cases}
\label{sigmaDynamics}	
\end{eqnarray}
The indoor temperature dynamics of the home is governed by\begin{eqnarray}
\dot{\theta}(t) = -\alpha\left(\theta(t) - \theta_{a}(t)\right) - \beta P \sigma(t),
\label{thetaDynamics}	
\end{eqnarray}
where the parameters $\alpha, \beta>0$ denote the heating time constant and thermal conductivity, respectively. Here, $P$ denotes the amount of thermal power drawn by the AC when it is ON ($\sigma(t)=1$). The electrical power drawn, $P_{e}$, is related to the thermal power $P$, via the formula $P_{e}=\frac{P}{\eta}$, where the parameter $\eta>0$ denotes the efficiency of the AC. 


\subsection{Direct Load Control Strategy}
In this paper, we propose an architecture where the setpoint of an individual AC will be changed over time by the LSE through a  velocity control strategy. The rate of change of setpoint $s(t)$ will constitute a ``control variable" that can be used by the LSE. This velocity command, suitably modulated so that the comfort zone is never violated, can be followed by an AC locally without the LSE being aware of the load's chosen values of $L_0,U_0$ or its current temperature $\theta(t)$. Yet, by squeezing the band close to the upper or lower limits, it can cause a certain fraction of loads to turn ON/OFF in response, thereby collectively invoking demand response. ACs that are ON with temperature trajectory headed downward can be intercepted by a rising lower boundary ($L_{t}$) to turn OFF. Similarly, ACs that are OFF with temperature trajectory headed upward can be intercepted by a falling upper boundary ($U_{t}$) and to force them turn ON. Thus, the demand response can induce less electricity consumption by issuing positive velocity commands, and it can also induce more cooling which allows storage of energy in the homes. This will be detailed in Section \ref{LoadControlSubsubsection}.

\begin{figure}[t]
  \centering
    \includegraphics[width=0.4\textwidth]{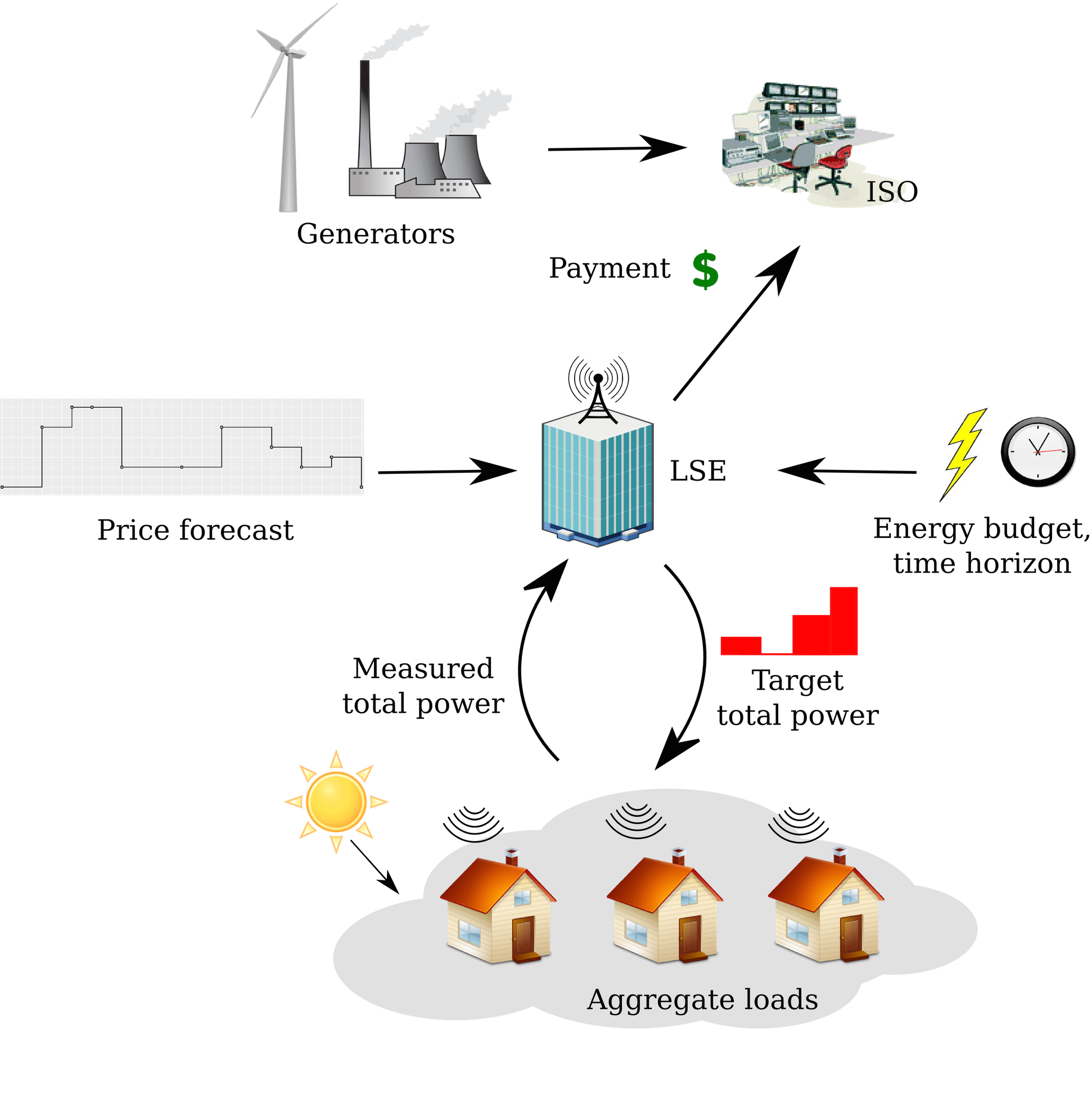}
    \vspace*{-0.2in}
    \caption{A schematic of the proposed architecture for thermal inertial load management by the LSE.}
    \label{HighLevelSchematic}
    \vspace*{-0.15in}
\end{figure}

\begin{figure*}[t]
  \centering
    \includegraphics[width=0.72\textwidth]{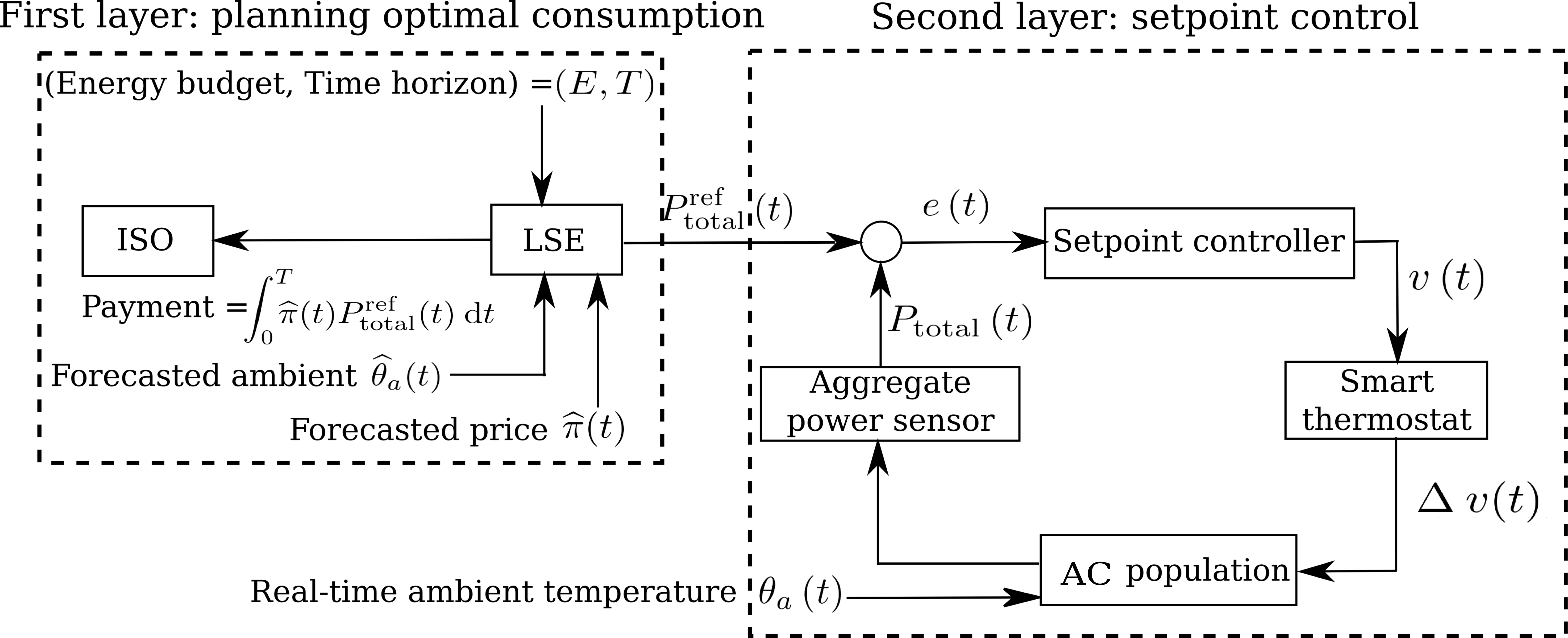}
    \caption{The two layer hierarchical strategy of the LSE.}
    \label{TwoLayerBlockDiagmHorizontal}
    \vspace*{-0.1in}
\end{figure*}

\subsection{Contracts}
If a home agrees to allow the LSE to manage the energy consumption of its AC, the LSE is obligated to respect privacy in the sense that it may influence the consumption but should do so without ever measuring the state $\{s(t), \theta(t), \sigma(t)\}$ of any individual AC. Furthermore, the LSE must guarantee that the indoor temperature $\theta(t)$ lies within the home's comfort range $[L_{0},U_{0}]$, at all times. These privacy and comfort range guarantees will constitute a contract between the LSE and an individual home. Larger the value of $\Delta$, greater is the flexibility of the load that an LSE can potentially exploit in shaping demand response. The value of the contract therefore depends on the magnitude of $\Delta$. A home may choose its tolerance $\Delta$, and is charged lesser for electricity the larger $\Delta$ it can tolerate. In Section \ref{ContractSectionLabel}, we will show how the LSE can determine the cost of providing electricity to a customer as a function of its flexibility $\Delta$, which it can then use to determine how to price contracts.


\subsection{Planning in the Day-Ahead Market} 
Much of the electricity purchase takes pace in the day-ahead market. In Section \ref{OperationalPlanningSubsection}, we will show how the LSE can operationally plan the amount of energy to purchase in each of the 24 one-hour periods of the day-ahead market,
so that it can support the loads at least cost. The goal of such planning is to ensure that the LSE purchases more energy at times when the day-ahead prices are low, and less energy in periods when day-ahead prices are high. The LSE must do so taking into consideration the flexibilities of the collection of homes and the need to keep their temperatures within their comfort zones. 

%


\section{Measurement and Control Architecture and Strategy for the LSE}
\label{ArchitectureStrategySection}
We propose a novel architecture (Fig. \ref{HighLevelSchematic}) and operational planning and control strategies for the LSE to 
manage a population of $N$ ACs over a finite time interval $[0,T]$, subject to the contractual obligations, privacy constraints, and load dynamics constraints, and to do so at low cost to the LSE. Our strategy consists of a planning phase where the LSE contracts to purchase the optimal amount of energy at low cost in the day-ahead market so that it can support the aggregate load of the homes under contract. 
We term this \emph{operational planning}. For $t\in[0,T]$, the LSE takes the measured real time total power consumption $P_{\text{total}}(t)$ as feedback to compute a control signal $v(t)$ that it broadcasts to all homes, requesting the setpoint of the $i$\textsuperscript{th} AC at time $t$, denoted as $s_{i}(t)$, be moved according to the \emph{velocity control} policy
 \begin{eqnarray}
 \dot{s}_{i}(t) = \Delta_{i} \: v(t), \quad i=1,\hdots,N.
 \label{SetpointVelocityControl}	
 \end{eqnarray}
Homes with different contractual $\Delta_{i}$ correspond to different comfort tolerances, and experience different setpoint velocities. The \emph{operational planning}, followed by \emph{setpoint velocity control}, comprise a two layer hierarchical strategy (Fig. \ref{TwoLayerBlockDiagmHorizontal}) for the LSE, which we expound next.

\begin{figure}[h]
  \centering
    \includegraphics[width=0.45\textwidth]{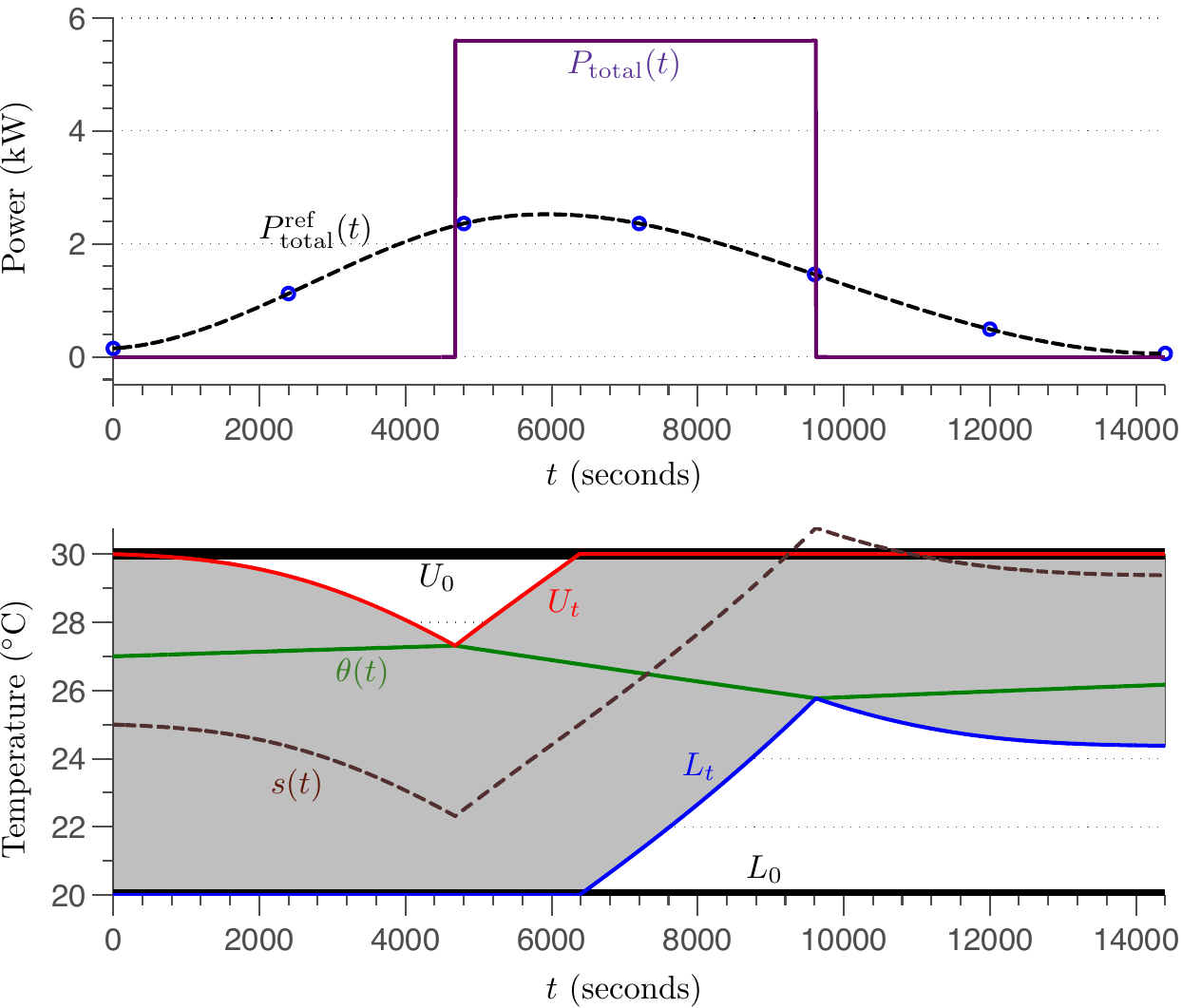}
    \caption{\emph{Top:} Using setpoint velocity control, the LSE makes the real time consumption $P_{\text{total}}(t)$ \emph{(solid line)}, shown here for a single AC with $\theta_{a}(t) = 32^{\circ}$C and $t\in[0,4\:\text{hours}]$, track the planned optimal reference consumption $P_{\text{total}}^{\text{ref}}(t)$ \emph{(dashed line)}. \emph{Bottom:} corresponding nonlinear deformation of the hysteretic band $[L_{0},U_{0}] \text{\emph{(thick black lines)}} \mapsto [L_{t},U_{t}]$ \emph{([blue line, red line])}, shown as the \emph{gray area}, in real time under the setpoint velocity control. Also shown are the $\theta(t)$ \emph{(solid green)} and $s(t)$ \emph{(dashed brown)} trajectories.}
    \label{ElasticDeadband}
\end{figure}

\subsubsection{Privacy Preserving Load Control}
\label{LoadControlSubsubsection} 

We suppose that the day-ahead operational planning, to be detailed in Section \ref{OperationalPlanningSubsection}, has led to an optimal reference aggregate power consumption trajectory $P_{\text{total}}^{\text{ref}}(t)$ for $0 \leq t \leq T$.
We now focus on the design of a real time privacy preserving setpoint \emph{velocity} controller for the LSE so as to make the actual total power consumption $P_{\text{total}}(t)$ track the planned optimal target consumption $P_{\text{total}}^{\text{ref}}(t)$. This corresponds to the ``second layer" in Fig. \ref{TwoLayerBlockDiagmHorizontal}. 
Assuming that the LSE can measure or estimate the aggregate output $P_{\text{total}}(t)$ in real time, define the error signal $e(t) := P_{\text{total}}(t) - P_{\text{total}}^{\text{ref}}(t)$. Then, a simple PID controller can generate a \emph{common} velocity command for all homes $v(t) := k_{p}e(t) + k_{i}\int_{0}^{t}e(\varsigma)\mathrm{d}\varsigma + k_{d}\frac{\mathrm{d}e}{\mathrm{d}t}$, with $k_{p}$, $k_{i}$, and $k_{d}$ being the proportional, integral, and derivative gains, respectively. We choose PID controller since it is easy to implement. Alternatively, one may use a model-based feedback controller for which the ``number density" \cite{GhaffariMouraKrstic2014DSMC} of ON/OFF ACs is approximated in real time. As a model-free implementation, the PID controller obviates this difficulty.

The LSE broadcasts the common velocity command $v(t)$ to all homes. Upon receiving the LSE's broadcast, an individual home's AC moves its setpoint according to (\ref{SetpointVelocityControl}). At any given time, a home with larger comfort range moves its setpoint faster than the one with smaller range, and hence provides more flexibility to the LSE for reference tracking. 
To respect the contractual comfort guarantee $L_{i0} \leq \theta_{i}(t) \leq U_{i0}$ for each $i=1,\hdots,N$, the setpoint controller (\ref{SetpointVelocityControl}), at any time $t$, utilizes the time-varying band $[L_{it}, U_{it}]$, where 
\begin{eqnarray}
L_{it} &:=& U_{i0} \wedge \left[L_{i0} \vee \left(s_{i}(t) - \Delta_{i}\right)\right],
\label{TimeVaryingLowerBoundary}	\\
U_{it} &:=& L_{i0} \vee \left[U_{i0} \wedge \left(s_{i}(t) + \Delta_{i}\right)\right].
\label{TimeVaryingUpperBoundary}
\end{eqnarray}
Thus, the width of the time-varying bands expand (contract) when the aggregate power consumption needs to be increased (reduced). In Fig. \ref{ElasticDeadband}, we show the deformation of the time-varying band $[L_{it}, U_{it}] \subseteq [L_{i0}, U_{i0}]$ for $N=1$. Furthermore, we notice that 
only the computation of $v(t)$ is centralized, but the computation (\ref{SetpointVelocityControl}), (\ref{TimeVaryingLowerBoundary}) and (\ref{TimeVaryingUpperBoundary}) are all decentralized.

From an architectural standpoint, there are two ways to interpret the computation in equations (\ref{SetpointVelocityControl})--(\ref{TimeVaryingUpperBoundary}). One may interpret them as part of control, and in that case the proposed real-time control algorithm becomes partially centralized (computation of $v(t)$) and partially decentralized (computation of (\ref{SetpointVelocityControl})--(\ref{TimeVaryingUpperBoundary})). This is unlike existing direct load control programs \cite{SouthernCaliforniaEdisonCA,XCelEnergyCOMNWI,EnergyWiseMD,PowerPartnerTX}, where the controls are fully centralized while the actuation are decentralized. Alternatively, one can consider only the centralized computation of $v(t)$ as control, and regard the decentralized computation of (\ref{SetpointVelocityControl})--(\ref{TimeVaryingUpperBoundary}) as part of smart actuation. In this interpretation, the proposed method differs from the existing direct load control programs in that traditional thermostatic actuation simply realize the control commands, whereas in our case, additional computation are performed at the actuators to modify the received control commands before realizing them.

\subsubsection{Estimating  the Total Power Consumption}
\label{SensingTotalPowerSubsubection}
We note first that the setpoint control loop is closed only over the aggregate total power $P_{\text{total}}(t)$ and not the individual homes' power draws, thus requiring no intrusive sensing of every home. The LSE only needs to measure or estimate $P_{\text{total}}(t)$ in real time, for the AC population it manages. 

\begin{figure}[t]
  \centering
    \includegraphics[width=0.37\textwidth]{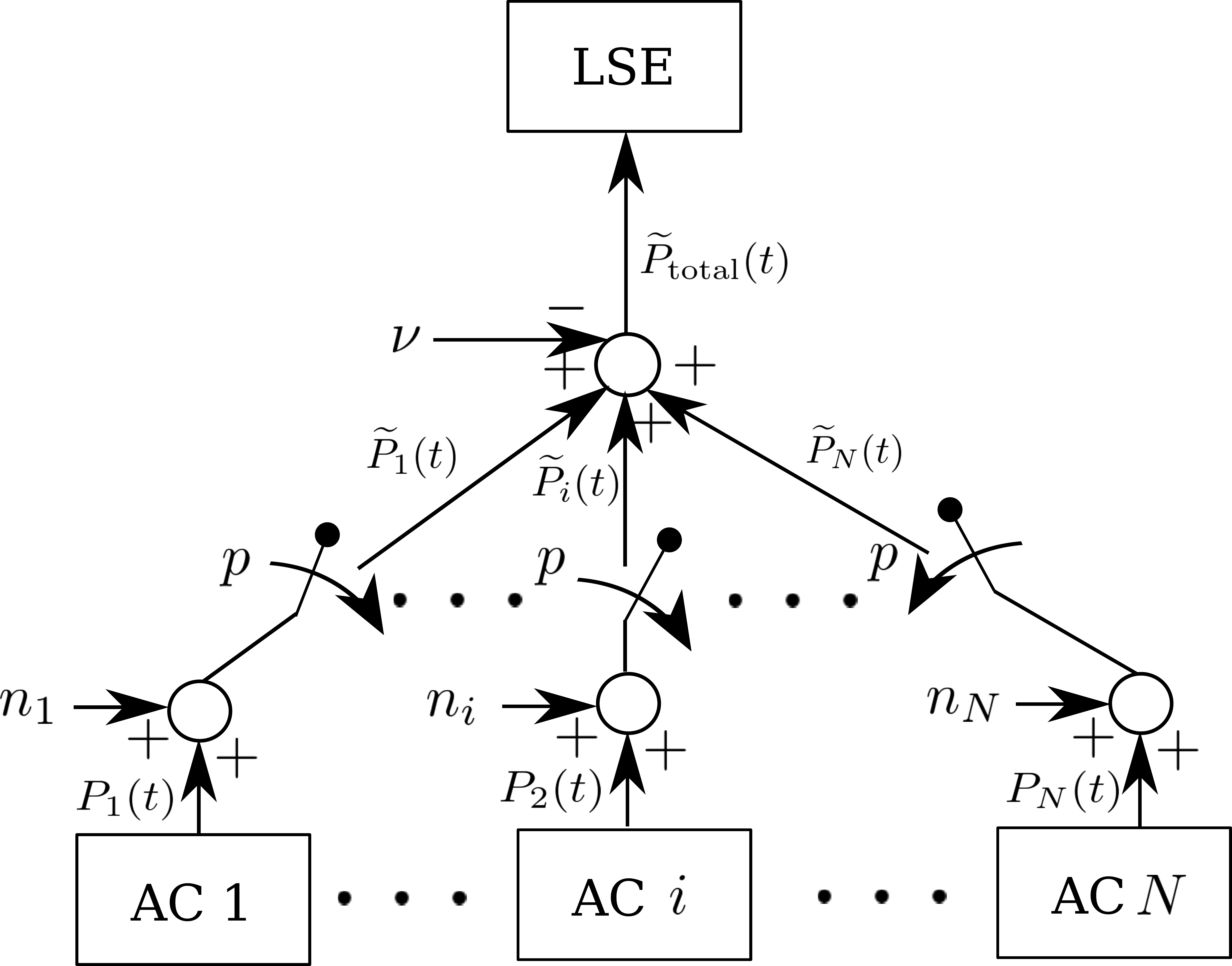}
    \caption{A schematic for the privacy preserving sensing architecture for the LSE to measure the real-time aggregate power $P_{\text{total}}(t)$ for controlling the setpoints of $N$ ACs it manages. Each AC, at any given time, consumes either zero or $P_{e}$ power. For small $\epsilon > 0$ and fixed $p\in(0,1)$, each home adds local i.i.d. noise from $\text{Gamma}(\frac{1}{pN},\frac{\epsilon}{pP_{e}})$, to the individual consumption at time $t$, and then transmits the noisy data with probability $p$. The sum of the reported consumptions, is first scaled by $\frac{1}{p}$, then perturbed further by subrtactive noise $\nu\sim\text{Exp}(\frac{\epsilon}{P_{e}})$, and the resulting value $\widetilde{P}_{\text{total}}(t)$ is sensed by the LSE. This guarantees (Appendix \ref{AppDiffPrivacy}) that at the LSE level, the individual consumption data remains $\epsilon$-differentially private, while the data transmitted by each AC remains secure against eavesdropping.}
    \label{DiffPrivacy}
    \vspace*{-0.15in}
\end{figure}

Using the technique of differential privacy \cite{DworkRoth2014}, we now show how this can be estimated even if the population of customers for the LSE constitutes only a subset of the total population whose consumption is being measured at a distribution substation. The $i$\textsuperscript{th} customer's home, with fixed probability $p$, reports a noise corrupted version $\widehat{P}_i(t) := P_i(t) + n_i$ of its power consumption $P_{i}(t)$ to the LSE at time $t$, where $n_i \sim \mbox{Gamma}(\frac{1}{p N}, \frac{\epsilon}{p P_e})$'s are chosen independently and identically distributed by the homes. The probability $p \in (0,1)$ satisfies $p N = \widehat{N}$, where $\widehat{N}$ is the number of customers' homes that report consumption to the LSE. From this, the LSE generates an estimate of total consumption for its population of customers $\widetilde{P}_{\text{total}}(t) = \widehat{P}_{\text{total}}(t) - \nu$, where $\widehat{P}_{\text{total}}(t) := \frac{N}{\widehat{N}}\sum_{i=1}^{\widehat{N}}\widehat{P}_i(t)$, and $\nu \sim \text{Exp}(\frac{\epsilon}{P_{e}})$. It can be shown (Appendix \ref{AppDiffPrivacy}) 
that $\widetilde{P}_{\text{total}}(t) := \frac{N}{\widehat{N}}\sum_{i=1}^{\widehat{N}} P_{i} + n$, where the additive zero mean noise random variable $n \sim \text{Lap}(\frac{P_{e}}{\epsilon})$, and the standard differential privacy formulation for the sum query ensures that an adversary would not be able to infer the $N\times 1$ consumption vector $\left(P_{1}(t), \hdots, P_{N}(t)\right)^{\top} \in \{0,P_{e}\}^{N}$, at any time $t$, if it were to snoop $\widetilde{P}_{\text{total}}(t)$. Also, even if $\widehat{P}_{i}(t)$ are eavesdropped, the same $\epsilon$-level of global differential privacy can be attained by this local implementation of differential privacy (different from the local differential privacy \cite{Warner1965,DuchiJordanWainright2013}), as shown in Fig. \ref{DiffPrivacy}. In Section \ref{CaseStudiesSectionLabel}, we will computationally analyze the tradeoff between privacy and tracking performance.


\subsubsection{Performance of the LSE's Strategy}
\label{PerformanceSubsection}
Since computing $P_{\text{total}}^{\text{ref}}(t)$ depends on both price and ambient temperature forecasts $\widehat{\pi}(t)$ and $\widehat{\theta}_{a}(t)$, the LSE's plan, and consequently the tracking performance during real time setpoint velocity control, are affected by the uncertainties in these forecasts, as well as the uncertainties in real time ambient temperature $\theta_{a}(t)$. Let $\Omega$ be the set of all tuples of forecasted price, forecasted and real time ambient temperatures. Then each forecast scenario $\omega\in\Omega$ refers to a triplet of correlated 
trajectories $(\widehat{\pi}(\omega,t),\widehat{\theta}_{a}(\omega,t), \theta_{a}(\omega,t))$, and engenders an effective width stochastic process for the $i$\textsuperscript{th} AC's band at time $t$, denoted hereafter as $w_{\text{eff}}\left(i,\omega,t\right)$, given by
\begin{eqnarray}
w_{\text{eff}}\left(i,\omega,t\right) := U_{it} - L_{it} = \Delta_{i}\left[2 - \bigg\lvert\int_{0}^{t}v(\omega,\varsigma)\mathrm{d}\varsigma\bigg\rvert\right]^{+},
\label{EffectiveDeadbandWidth}	
\end{eqnarray}
where the last equality results from substituting (\ref{TimeVaryingLowerBoundary}) and (\ref{TimeVaryingUpperBoundary}) for $L_{it}$ and $U_{it}$ respectively, followed by some algebra (Appendix \ref{SomeAlgebra}). To statistically quantify the performance of the LSE's control strategy, we observe that the limit of control performance for the $i$\textsuperscript{th} AC, is 
\begin{eqnarray}
\xi_{T}\left(\omega\right) := \displaystyle\frac{1}{T}\displaystyle\lim_{\epsilon \downarrow 0} \displaystyle\frac{1}{\epsilon} \displaystyle\int_{0}^{T} \mathds{1}_{\{0 \leq w_{\text{eff}}(i,\omega,t) < \epsilon\}} \: \mathrm{d}t,
\label{NormalizedLocalTime}	
\end{eqnarray}
which is the normalized local time \cite{MarcusRosen} for the effective width stochastic process. In words, the random variable $\xi_{T}\left(\omega\right)$ measures the fraction of time $i$\textsuperscript{th} AC remains inflexible over the fixed horizon $[0,T]$, for a scenario $\omega \in \Omega$. Notice that (\ref{NormalizedLocalTime}) does not depend on $i$, which is a consequence of (\ref{SetpointVelocityControl}), \emph{i.e.}, zero effective width epochs are synchronized among all ACs. In Section \ref{CaseStudiesSectionLabel}, we will compute $\xi_{T}(\omega)$ for different forecast scenarios of ($\widehat{\pi}(\omega,t),\widehat{\theta}_{a}(\omega,t),\theta_{a}(\omega,t)$), available from the historical data.

\subsubsection{Assumptions}
We make the following assumptions.
\begin{itemize}
\item The ambient temperature trajectories $\widehat{\theta}_{a}(t)$, $\theta_{a}(t)$, and the price forecast $\widehat{\pi}(t)$ are positive smooth (continuously differentiable) functions of time $t$.

\item ACs are necessary, \emph{i.e.}, for all $t \in[0,T]$, we have $\widehat{\theta}_{a}(t),\theta_{a}(t) > \max_{i=1,\hdots,N}U_{i0}$.

\item Without loss of generality, the initial indoor temperatures $\theta_{i0} := \theta_{i}(0) \in [L_{i0},U_{i0}]$, for all $i=1,\hdots,N$.   
	
\end{itemize}

\subsubsection{Operational Planning}
\label{OperationalPlanningSubsection}
We now address how the reference trajectory $P_{\text{total}}^{\text{ref}}(t)$ can be chosen so as to minimize the cost to the LSE. We suppose that the LSE is
exposed to a price forecast $\widehat{\pi}(t)$ over $[0, T]$ from a day-ahead market. Suppose that the LSE serves $N$ homes with comfort tolerances $\{\Delta_{i}\}_{i=1}^{N}$. Further, for simplicity of exposition, we assume that each AC in the population, when ON, draws same thermal power $P$. The homes may have different thermal coefficients ($\alpha, \beta$), and the LSE only knows a joint distribution $\rho_{\alpha,\beta}$ across the population of its customers. To provision an optimal aggregate consumption profile $P_{\text{total}}^{\text{ref}}(t)$ over the time window $[0,T]$, the LSE makes its decision based on a price forecast $\widehat{\pi}(t)$ available from the energy market, energy budget $E$ available from the load forecast, and predicted ambient temperature trajectory $\widehat{\theta}_{a}(t)$ available from the weather forecast. This is depicted as the ``first layer" in Fig. \ref{TwoLayerBlockDiagmHorizontal}. 

For $i=1,\hdots,N$, let the indicator variable $u_{i}(t) \in \{0,1\}$ denote whether the $i$\textsuperscript{th} AC is ON/OFF at time $t$. Let $n_{\text{ON}}(t)$ denote the number of ACs that are ON at time $t$. Since the aggregate electrical power drawn by the AC population at time $t$ is $P_{e}(t) = \frac{P}{\eta}n_{\text{ON}}(t) = \frac{P}{\eta}\sum_{i=1}^{N}u_{i}(t)$, and the procurement cost for total energy consumption over $[0,T]$ is $\int_{0}^{T}P_{\text{total}}^{\text{ref}}(t)\widehat{\pi}(t)\mathrm{d}t$, the LSE's objective is to
\begin{eqnarray}
\underset{u_{1}(t),\hdots,u_{N}(t)\in\{0,1\}^{N}} {\text{minimize}}	\displaystyle\frac{P}{\eta} \displaystyle\int_{0}^{T} \widehat{\pi}(t) \displaystyle\sum_{i=1}^{N}u_{i}(t) \: \mathrm{d}t, \label{PlanningCostFcn}
\end{eqnarray}
subject to the constraints 
\begin{eqnarray}
			&&\dot{\theta}_{i}(t) = -\alpha_{i}\left(\theta_{i}(t) - \widehat{\theta}_{a}(t)\right) - \beta_{i}Pu_{i}(t), \label{DynamicsConstraint}\\
&&\displaystyle\frac{P}{\eta}\displaystyle\int_{0}^{T}\displaystyle\sum_{i=1}^{N}u_{i}(t)\:\mathrm{d}t = E, \label{EnergyBudgetConstraint}\\
&&L_{i0} \leq \theta_{i}(t) \leq U_{i0},\label{ComfortRangeConstraint}
\end{eqnarray}
where $[L_{i0},U_{i0}] := [s_{i0} - \Delta_{i}, s_{i0} + \Delta_{i}]$, and $i = 1,\hdots,N$. We notice that the dynamics constraints (\ref{DynamicsConstraint}) and the comfort range constraints (\ref{ComfortRangeConstraint}) are decoupled, while the cost function (\ref{PlanningCostFcn}) and the energy budget constraint (\ref{EnergyBudgetConstraint}) are coupled. The constraint (\ref{EnergyBudgetConstraint}) provides a modeling flexibility for the LSE, in case a total energy budget $E>0$ is available from load forecast.

Equations (\ref{PlanningCostFcn})--(\ref{ComfortRangeConstraint}) represent a continuous time deterministic optimal control problem. Denoting its solution as $\{u_{i}^{*}(t)\}_{i=1}^{N}$, expressed as an open-loop optimal control, the optimal reference consumption is given by $P_{\text{total}}^{\text{ref}}(t) = \frac{P}{\eta}\sum_{i=1}^{N}u_{i}^{*}(t)$. In Section \ref{CaseStudiesSectionLabel}, we will follow a ``discretize-then-optimize" approach to numerically solve (\ref{PlanningCostFcn})--(\ref{ComfortRangeConstraint}).

\begin{figure}[t]
  \centering
    \includegraphics[width=0.49\textwidth]{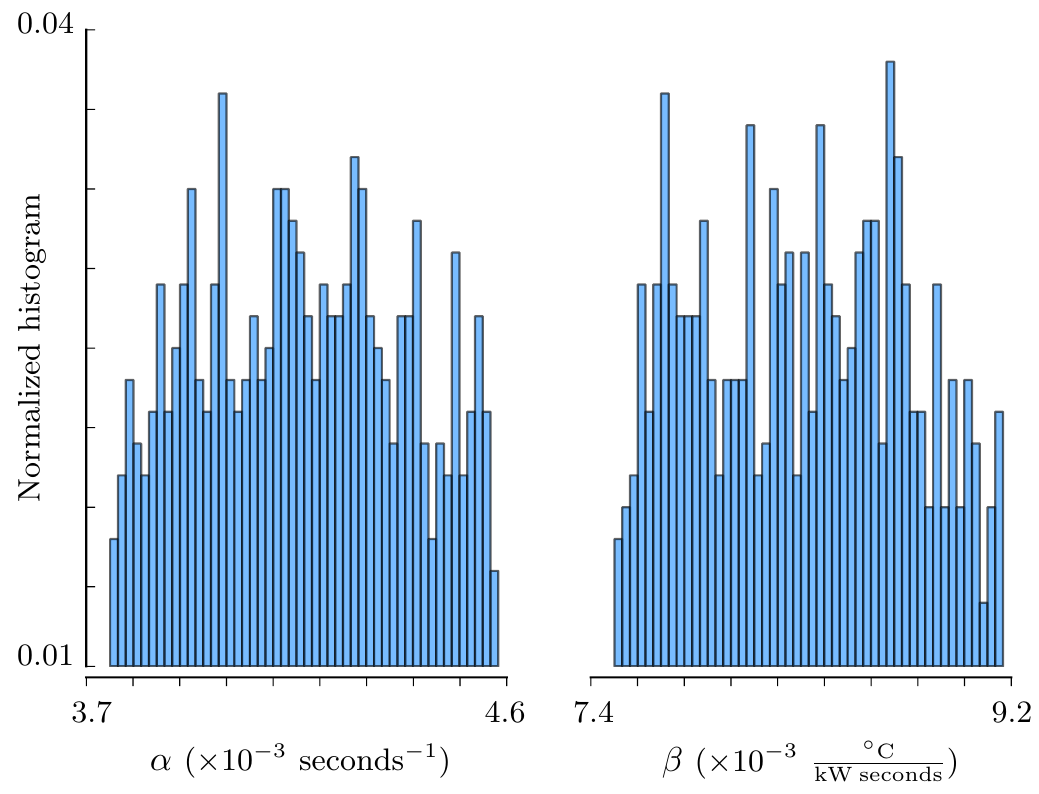}
    \caption{Normalized histograms for $\alpha \sim \mathcal{N}\left(\mu_{\alpha},0.1\mu_{\alpha},0.9\mu_{\alpha},1.1\mu_{\alpha}\right)$ (\emph{left}), and for $\beta \sim \mathcal{N}\left(\mu_{\beta},0.1\mu_{\beta},0.9\mu_{\beta},1.1\mu_{\beta}\right)$ (\emph{right}), with $\mu_{\alpha} = \frac{1}{RC}$ h\textsuperscript{-1}, $\mu_{\beta} = \frac{1}{C}$ $^{\circ}$C/kWh, used to generate the heterogeneous population of thermal inertial loads.}
    \label{AlphaBetaDistribution}
    \vspace*{-0.1in}
\end{figure}
\begin{figure}[t]
  \centering
  \hspace*{-0.2in}
    \includegraphics[width=0.5\textwidth]{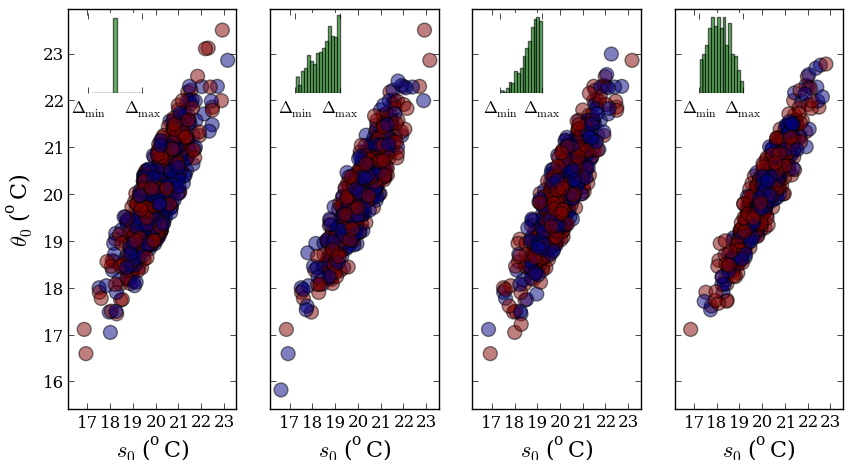}
    \caption{Initial conditions for $N=500$ homes, for four different contractual comfort tolerance ($\Delta$) distributions, shown \emph{left} to \emph{right}. The \emph{red} circles denote ON ($\sigma_{0}=1$), and the \emph{blue} circles denote OFF ($\sigma_{0}=0$) initial states.}
    \label{InitialConditions}
\vspace*{-0.1in}    
\end{figure}

\begin{figure*}[htpb]
  \centering
    \includegraphics[width=\textwidth]{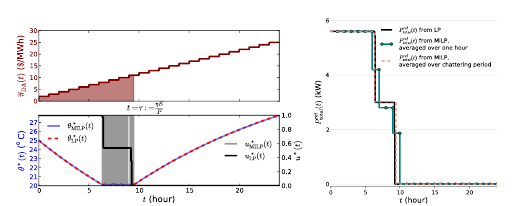}
    \caption{\emph{Left top:} Increasing day-ahead price forecast $\widehat{\pi}_{\text{DA}}(t)$ as a piecewise constant function for each hour, with $\tau := \frac{\eta E}{P}$ shown; \emph{left bottom}: the corresponding optimal indoor temperature $\theta^{*}(t)$, and ON/OFF control $u^{*}(t)$ obtained by first discretizing (\ref{PlanningCostFcn})--(\ref{ComfortRangeConstraint}) for a single AC, and then solving the resulting MILP (using Gurobi) and its LP relaxation (using MATLAB). In this simulation, $[L_{0},U_{0}] = [20^{\circ}\mathrm{C}, 30^{\circ}\mathrm{C}]$, the ambient forecast is constant, \emph{i.e.}, $\widehat{\theta}_{a}(t) \equiv 32^{\circ}\mathrm{C}$, and the parameters $\alpha, \beta, P, \eta$ are from Table 1, p. 1392 in \cite{Callaway2009Energy}. \emph{Right:} The optimal consumption trajectory $P_{\text{total}}^{\text{ref}}(t)$ computed from the MILP and its LP relaxation, for the planning problem shown in the \emph{left}.}
    \label{MILPvsLP}
\end{figure*}

\subsubsection{Feasibility of the Planning Problem}
\label{FeasibilitySubsubsection}
Let $\tau := \frac{\eta E}{P}$, and notice that constraint (\ref{EnergyBudgetConstraint}) imposes a necessary feasibility condition
\begin{eqnarray}
0 \leq \overline{\tau} := \frac{\tau}{NT} = \frac{\eta E}{NPT} \leq 1.
\label{Feasibility}	
\end{eqnarray}
Given an ambient forecast $\widehat{\theta}_{a}(t)$ and parameters of the TCL population, to respect constraint (\ref{ComfortRangeConstraint}), $\overline{\tau}$ is restricted to
\begin{eqnarray}
0 \leq \overline{\tau}_{\ell} \leq \overline{\tau} \leq \overline{\tau}_{u} \leq 1,
\label{FeasibilitywZeroDynamics}	
\end{eqnarray}
where $\overline{\tau}_{\ell} := \frac{\tau_{\ell}}{NT} = \frac{\eta E_{\ell}}{NPT}$, and $\overline{\tau}_{u} = \frac{\eta E_{u}}{NPT}$. Here, $E_{\ell}$ (resp. $E_{u}$) is the aggregate energy consumed if the indoor temperatures for each home in the population were to be restricted at their private upper (resp. lower) setpoint boundaries, thus resulting in the lowest (resp. highest) total energy consumption while respecting (\ref{ComfortRangeConstraint}). In other words, $E_{\ell} = \frac{P}{\eta}\sum_{i=1}^{N}\int_{0}^{T}u_{i}(t)\:\mathrm{d}t$, where the corresponding controls are $u_{i}(t) = \frac{\alpha_{i}}{\beta_{i}P}(\widehat{\theta}_{a}(t) - U_{i0})$, and hence $\overline{\tau}_{\ell} = \frac{1}{NP}(\sum_{i=1}^{N}\frac{\alpha_{i}}{\beta_{i}}(\langle\widehat{\theta}_{a}\rangle - U_{i0}))$,
where $\langle\widehat{\theta}_{a}\rangle := \frac{1}{T}\int_{0}^{T}\widehat{\theta}_{a}(t)\:\mathrm{d}t$. Similar calculation yields $\overline{\tau}_{u} = \frac{1}{NP}(\sum_{i=1}^{N}\frac{\alpha_{i}}{\beta_{i}}(\langle\widehat{\theta}_{a}\rangle - L_{i0}))$. Equation (\ref{FeasibilitywZeroDynamics}) gives the necessary and sufficient conditions for feasibility of the planning problem (\ref{PlanningCostFcn})--(\ref{ComfortRangeConstraint}). Notice that if the constraint (\ref{EnergyBudgetConstraint}) is inactive, then $\frac{P}{\eta}\int_{0}^{T}\sum_{i=1}^{N}u_{i}^{*}\:\mathrm{d}t = E_{\ell}$. Also, note that (\ref{FeasibilitywZeroDynamics}) is equivalent to the energy inequality $E_{\min} \leq E_{\ell} \leq E \leq E_{u} \leq E_{\max}$, where $E_{\min} := 0$ and $E_{\max} := \frac{NPT}{\eta}$. For privacy preserving planning purpose, as earlier, the LSE can use noisy estimates and random samples for the initial conditions of the AC states $(s_{i0},\theta_{i0}, \sigma_{i0})_{i=1}^{N}$. Similarly, the thermal coefficient parameters $(\alpha_{i},\beta_{i})_{i=1}^{N}$ can be sampled from $\rho_{\alpha,\beta}$.

\section{Pricing the Contracts}\label{ContractSectionLabel}
To address the issue of how the LSE can estimate the cost of a contract for an individual home with a given flexibility $\Delta$, we employ a sensitivity analysis approach. The LSE can simply calculate the cost (\ref{PlanningCostFcn}) with and without an additional customer of the given type. 
Assuming the LSE has an existing customer population of $N$ ACs, let the optimal value of (\ref{PlanningCostFcn}) be $J^{*} := \int_{0}^{T}\widehat{\pi}(t) P_{\text{total}}^{\text{ref}}(t)\:\mathrm{d}t$. Recall that $\widehat{\pi}(t)$ has unit \$/MWh, and for $T=24$ hours, $J^{*}$ (in \$) denotes the daily energy procurement cost. For $i=1,\hdots,N$, let $J_{-i}^{*}$ denote the optimal value of (\ref{PlanningCostFcn}) with the $i$\textsuperscript{th} AC removed from the population. Clearly, $J_{-i}^{*} > J^{*}$, and the increase in per day cost $J_{-i}^{*} - J^{*}$ represents the marginal value (\$/day) of an AC with comfort tolerance $\Delta_{i}$. Thus, the LSE can use the graph of $J_{-i}^{*} - J^{*}$ as function of $\Delta_{i}$, as a price chart for the thermal inertial load management service. When  an individual AC, as new customer to the LSE, specifies its choice of comfort tolerance, this graph will determine the price of the new contract between the LSE and the AC. This will be illustrated further in Section \ref{CaseStudiesSectionLabel}.

\section{Numerical Results}\label{CaseStudiesSectionLabel}


\subsection{Parameters and Simulation Setup}
\label{ParameterSubsubsection}

\begin{figure}[htpb]
  \centering
    \includegraphics[width=0.49\textwidth]{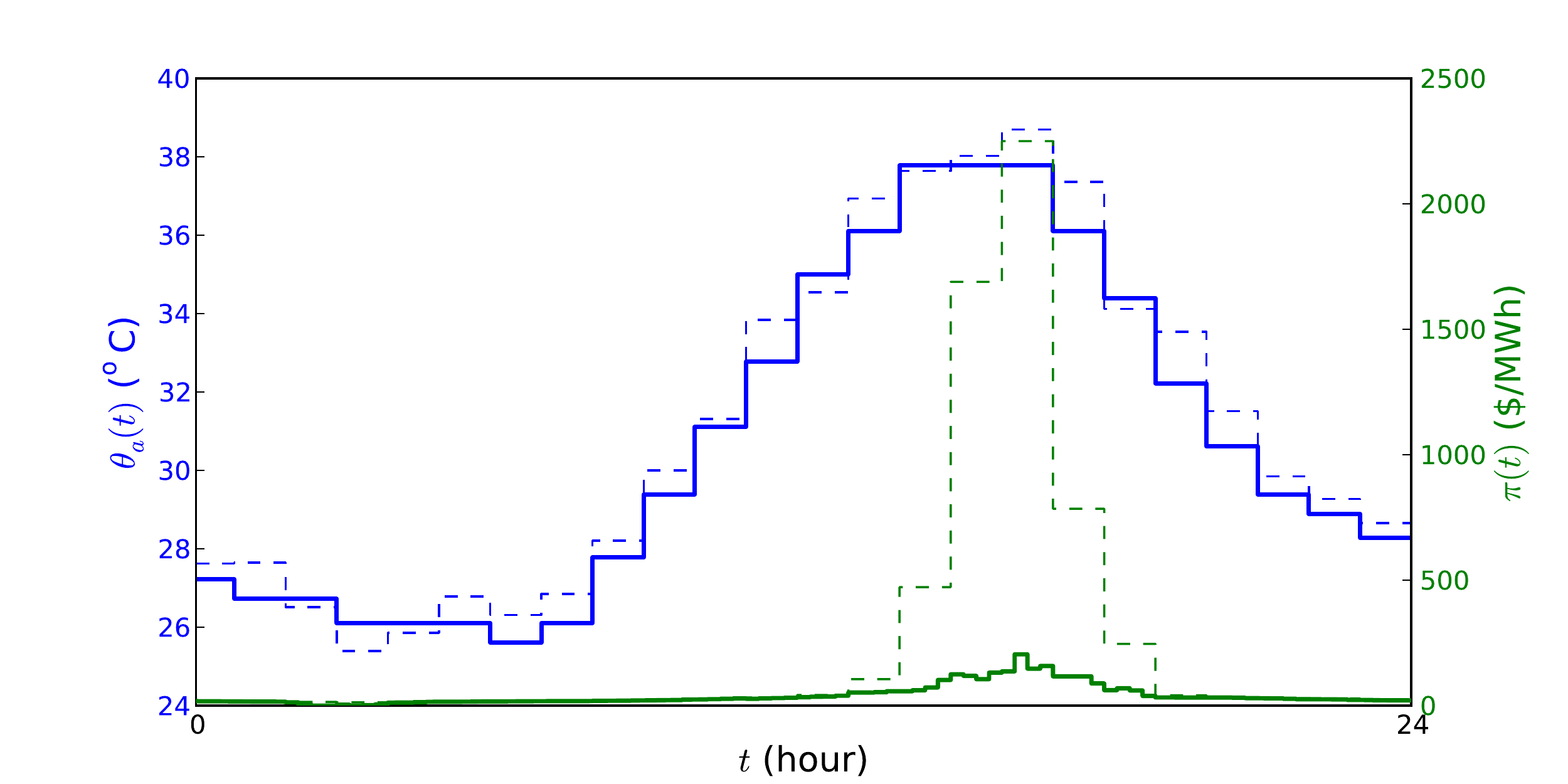}
    \caption{The day-ahead forecasted ambient temperature $\widehat{\theta}_{a}(t)$ (\emph{dashed blue}), real-time ambient temperature $\theta_{a}(t)$ (\emph{solid blue}), the day-ahead forecasted price of energy $\widehat{\pi}_{\text{DA}}(t)$ (\emph{dashed green}), and real-time price of energy $\pi_{\text{RT}}(t)$ (\emph{solid green}) data used in Section \ref{HomeCaseStudySectionLabel}.} 
    \label{PriceAmbient}
\end{figure}

\begin{figure*}[t]
  \centering
  \hspace*{-0.14in}
    \includegraphics[width=1.05\textwidth]{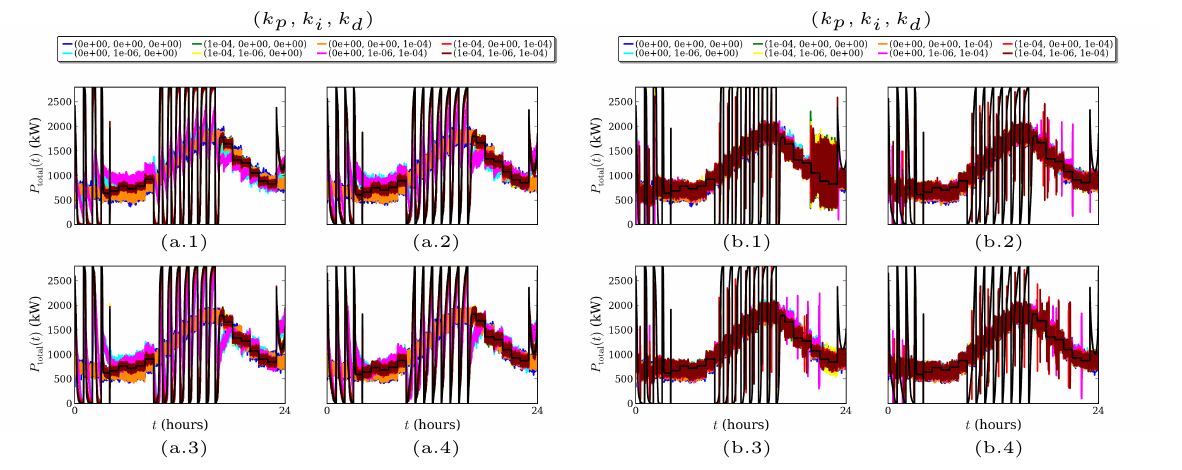}
    \caption{(a) The setpoint tracking performance for the PID velocity controller implemented \emph{without} (\ref{TimeVaryingLowerBoundary}) and (\ref{TimeVaryingUpperBoundary}); and (b) the same \emph{with} (\ref{TimeVaryingLowerBoundary}) and (\ref{TimeVaryingUpperBoundary}), for different combinations of PID gain tuple $(k_{p},k_{i},k_{d})$, shown as different colors in the legend box at the top. The four subfigures, denoted as (a.1)--(a.4) and (b.1)--(b.4), refer to four different contractual $\Delta$ distributions mentioned in Section \ref{ParameterSubsubsection}.}
    \label{Prealtime}
    \vspace*{-0.1in}
\end{figure*}

Following Table 1 (p. 1392) in \cite{Callaway2009Energy}, we fix the thermal resistance $R = 2^{\circ}$C/kW, the thermal capacitance $C = 10$ kWh/$^{\circ}$C, the thermal power drawn by an ON AC as $P = 14$ kW, and the load efficiency $\eta = 2.5$. We choose the distribution of thermal coefficient parameters as truncated Gaussians (Fig. \ref{AlphaBetaDistribution}), 
given by $\alpha \sim \mathcal{N}\left(\mu_{\alpha},0.1\mu_{\alpha},0.9\mu_{\alpha},1.1\mu_{\alpha}\right)$ 
and $\beta \sim \mathcal{N}\left(\mu_{\beta},0.1\mu_{\beta},0.9\mu_{\beta},1.1\mu_{\beta}\right)$, 
with $\mu_{\alpha} = \frac{1}{RC}$ h\textsuperscript{-1}, $\mu_{\beta} = \frac{1}{C}$ $^{\circ}$C/kWh. We consider a population of $N = 500$ homes, a time horizon of $T = 24$ hours, and suppose that the consumers have comfort tolerances ($\Delta$) between $\Delta_{\min} = 0.1^{\circ}$C and $\Delta_{\max} = 1.1^{\circ}$C. We investigate four cases: (i) the entire population of customers has identical $\Delta \equiv 1^{\circ}$C, (ii) the population has  i.i.d. uniform $\Delta$ over $[\Delta_{\min},\Delta_{\max}]$, (iii) the population has i.i.d. right triangular $\Delta$ distribution over $[\Delta_{\min},\Delta_{\max}]$ with peak at $\Delta_{\max}$, (iv) same as (iii) but with peak at $\Delta_{\min}$. Intuitively, in case (iii) (case (iv)), there are more homes with large (small) $\Delta$, and the population offers higher (lower) flexibility for the LSE's control strategy.

We sample the initial conditions $(s_{i0},\theta_{i0})$ from a bivariate Gaussian $\mathcal{N}\left(\mu_{0},\Sigma_{0}\right)$, with mean vector $\mu_{0} = (20, 20)^{\top}$ and covariance matrix $\Sigma_{0} = \begin{bmatrix}1 & 0.5\\0.5 & 3\end{bmatrix}$, subject to the constraint $s_{i0} - \Delta_{i} \leq \theta_{i0} \leq s_{i0} + \Delta_{i}$, where the comfort tolerance samples $\Delta_{i}$ are drawn from distributions according to case (i)-(iv), and $i=1,\hdots,N$. For each tuple $(s_{i0},\theta_{i0})$ thus generated, we assign $\sigma_{i0} = 1$ or $0$ with probability $0.5$. The resulting initial conditions are shown in Fig. \ref{InitialConditions}.

\subsection{Case Studies} 
\label{HomeCaseStudySectionLabel}

\subsubsection{Direct Numerical Solution of the Planning Problem}
\label{MILPvsLPsubsubsection}
To elucidate the numerical solution of (\ref{PlanningCostFcn})--(\ref{ComfortRangeConstraint}), let us consider $N=1$ home with $[L_{0},U_{0}] = [20^{\circ}\mathrm{C}, 30^{\circ}\mathrm{C}]$, $\theta_{0} = 25^{\circ}\text{C}$, constant ambient temperature forecast $\widehat{\theta}_{a}(t) = 32^{\circ}\mathrm{C}$, day-ahead price forecast $\widehat{\pi}_{\text{DA}}(t)$ and $\tau$ as shown in Fig. \ref{MILPvsLP}, left top. We discretize (\ref{PlanningCostFcn})--(\ref{ComfortRangeConstraint}) with $\Delta t = 1$ minute time step size, and use Euler discretization for the ODE (\ref{DynamicsConstraint}), to transcribe the optimal control problem (\ref{PlanningCostFcn})--(\ref{ComfortRangeConstraint}) as a mixed integer linear programming (MILP) problem in decision vector $\{\theta(1), u(1), \hdots, \theta(T), u(T)\}$. In Fig. \ref{MILPvsLP} left bottom and right, we compare the MILP solution, computed using Gurobi \cite{Gurobi2015}, with the solution of its linear programming (LP) relaxation, computed using MATLAB. Intuitively, since $\widehat{\pi}_{\text{DA}}(t)$ is increasing, it is optimal to stay ON during $[0,\tau]$, and OFF thereafter. We observe that $\left(\theta^{*}_{\text{MILP}}(t),u^{*}_{\text{MILP}}(t)\right)$ matches with $\left(\theta^{*}_{\text{LP}}(t),u^{*}_{\text{LP}}(t)\right)$ except at times $t \in [t_{\text{hit}},\tau]$, where $t_{\text{hit}}$ is the first hitting time to $L_{0}$. In particular, during $[t_{\text{hit}},\tau]$, the solution $\theta^{*}_{\text{MILP}}(t)$ (resp. $\theta^{*}_{\text{LP}}(t)$) chatters (resp. slides) along $L_{0}$. In the chattering period, $u^{*}_{\text{MILP}}(t)$ switches between 0 and 1 at every time step, and as shown in Fig. \ref{MILPvsLP} right, $u^{*}_{\text{MILP}}(t) \in \{0,1\}$ averaged over the chattering period yields $u^{*}_{\text{LP}}(t)\in [0,1]$, \emph{i.e.}, the proper fractional values of $u^{*}_{\text{LP}}(t)$ can be interpreted as the fraction of time the AC remains ON during $[t_{\text{hit}},\tau]$.

For $N$ homes with 1 minute time step size, the planning problem requires solving an MILP in $24 \times 60 \times N \times 2$ variables, and in our numerical experiments with Gurobi, solving the MILP even for $N=2$ was found to have more than 24 hours of CPU runtime. Armed with the physical meaning of the LP relaxation discussed above, in this paper, we solve (\ref{PlanningCostFcn})--(\ref{ComfortRangeConstraint}) under control convexification $\left(u_{1}(t), \hdots, u_{N}(t)\right) \in \{0,1\}^{N} \mapsto \left(u_{1}(t), \hdots, u_{N}(t)\right) \in [0,1]^{N}$. This LP relaxation allows us to compute for large $N$ (e.g. $N=500$ in 10 minutes CPU time of which only 36 seconds are needed to run the LP solver), which is of interest from the LSE's perspective.

\subsubsection{Planning and Setpoint Control for Actual Data}
\label{ActualDataSubsubsection}
We now apply the LSE's strategy proposed in this paper, to the day-ahead price forecast $\widehat{\pi}_{\text{DA}}(t)$ data for Houston on August 10, 2015, available \cite{DApriceERCOT} from ERCOT. Further, we use Houston weather station data, for day-ahead forecasted ambient temperature $\widehat{\theta}_{a}(t)$ on August 10, 2015, and for real time ambient temperature $\theta_{a}(t)$ on August 11, 2015. With these price and ambient temperature data (Fig. \ref{PriceAmbient}), and using the parameter values in Section \ref{ParameterSubsubsection}, we first solve the day-ahead planning problem described in Section \ref{OperationalPlanningSubsection} to design the optimal aggregate consumption, and then track that reference consumption in real time via setpoint velocity control described in Section \ref{LoadControlSubsubsection}. As in Section \ref{MILPvsLPsubsubsection}, the planning problem is solved with time step size equal to $1$ minute, and with $\overline{\tau}=\frac{1}{3}$, which is feasible from (\ref{FeasibilitywZeroDynamics}). The setpoint velocity control is implemented with a smaller time step size of $1$ second, to emulate the actual dynamics possibly being different from the model used in planning computation.

\begin{table}[t]
\caption{Energies (in MWh) mentioned in Section \ref{FeasibilitySubsubsection}, associated with the optimal planning problem in Section \ref{ActualDataSubsubsection}. The gray columns show the real time controlled ($E_{\mathrm{c}}$) and uncontrolled ($E_{\mathrm{unc}}$) energy consumptions in MWh.}
\label{EnergyTable}

\centering
\resizebox{\columnwidth}{!}{%
\begin{tabular}{|c|c|c|c|>{\columncolor[gray]{0.8}}c|>{\columncolor[gray]{0.8}}c|c|c|}

\hline

$\Delta$ cases & $E_{\min}$ & $E_{\ell}$ & $E$ & $E_{\mathrm{c}}$ & $E_{\mathrm{unc}}$ & $E_{u}$ & $E_{\max}$\\ \hline\hline
\begin{tikzpicture}[baseline=0]
    \draw (-0.1,-0.1) rectangle (-0.1,0.3);
    \draw (-0.1,-0.2) node[]{\tiny{$1.0$}}; \addvmargin{1mm}\end{tikzpicture} & \multirow{12}{*}{0} & 21.5088 & \multirow{12}{*}{22.4} & 22.4681 & 25.6821 & 26.3328 & \multirow{12}{*}{67.2}\\ \cline{1-1}\cline{3-3}\cline{5-6}
\begin{tikzpicture}[baseline=0]\draw (0,0) node[anchor=north]{\tiny{$0.1$}}
  -- (0.5,0) node[anchor=north]{\tiny{$1.1$}}
  -- (0.5,0.5) node[anchor=south]{}
  -- (0,0.5) node[anchor=south]{}
  -- cycle; \addvmargin{1mm}\end{tikzpicture} & & 22.1496 & & 22.4485 & 25.6905 & 26.6778 & \\
\cline{1-1}\cline{3-3}\cline{5-6}      
\begin{tikzpicture}[baseline=0]\draw (0,0) node[anchor=north]{\tiny{$0.1$}}
  -- (0.5,0) node[anchor=north]{\tiny{$1.1$}}
  -- (0.5,0.5) node[anchor=south]{}
  -- cycle; \addvmargin{1mm}\end{tikzpicture} & & 21.9287 & & 22.4690 & 25.7193 & 25.9663 & \\ 
  \cline{1-1}\cline{3-3}\cline{5-6}
\begin{tikzpicture}[baseline=0]\draw (0,0) node[anchor=north]{\tiny{$0.1$}}
  -- (0.5,0) node[anchor=north]{\tiny{$1.1$}}
  -- (0,0.5) node[anchor=south]{}
  -- cycle; \addvmargin{1mm}\end{tikzpicture} & & 22.5625 & & 22.4299 & 25.0748 & 25.2126 & \\ \cline{1-1}\cline{3-3}\cline{5-6}\hline				
\end{tabular}
}
\vspace*{-0.1in}
\end{table}

\begin{figure}[t]
  \centering
    \includegraphics[width=0.49\textwidth]{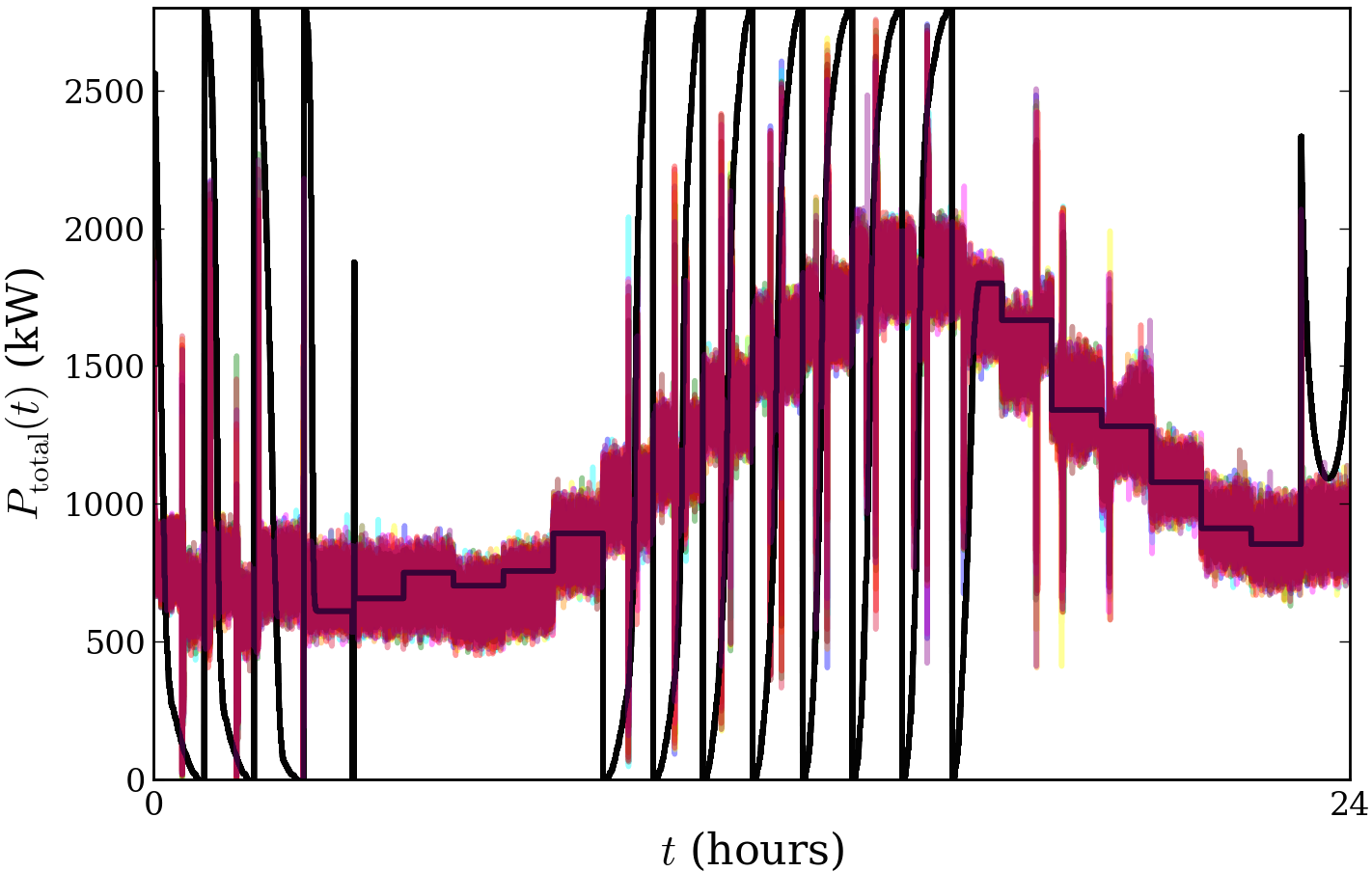}
    \caption{Tracking performance of the PID setpoint velocity controller with local implementation of differential privacy, as described in Section \ref{SensingTotalPowerSubsubection}, with $\epsilon = 0.1$, $p=0.9$, and $(k_{p},k_{i},k_{d}) = (10^{-4}, 10^{-6}, 10^{-4})$. The black curve above is the same as that in Fig. \ref{Prealtime}(b.2), \emph{i.e.}, $P_{\text{total}}^{\text{ref}}(t)$ for $\Delta$ uniformly distributed in $[0.1,1.1] ^{\circ}$C. Different colored $P_{\text{total}}(t)$ correspond to 10 different realizations of the noise random variable $n$.}
    \label{NoisyTracking}
   \vspace*{-0.1in}
\end{figure}

In Fig. \ref{Prealtime}, two sets of four subfigures, labeled as (a.1)--(a.4) and (b.1)--(b.4), are shown. The subfigures (a.1)--(a.4) (resp. (b.1)--(b.4)), show the tracking performance of the setpoint velocity controller (\ref{SetpointVelocityControl}) in the absence of noise $n$, without (resp. with) enforcing  the comfort range constraints (\ref{ComfortRangeConstraint}) in real time via (\ref{TimeVaryingLowerBoundary}) and (\ref{TimeVaryingUpperBoundary}). The four subfigures in each set correspond to the four contractual cases mentioned in Section \ref{ParameterSubsubsection}. The black curve in each subfigure denotes $P_{\text{total}}^{\text{ref}}(t)$ obtained by solving the discretized LP for the control convexified (see Section \ref{MILPvsLPsubsubsection}) version of (\ref{PlanningCostFcn})--(\ref{ComfortRangeConstraint}) with the corresponding initial conditions and parameters. As shown in Table \ref{EnergyTable}, for each contractual case, $E:=\int_{0}^{T}P_{\text{total}}^{\text{ref}}(t)\:\mathrm{d}t$ satisfies the energy inequality mentioned in Section \ref{FeasibilitySubsubsection}. The gray columns in Table \ref{EnergyTable} show the actual consumptions $\int_{0}^{T}P_{\text{total}}(t)\:\mathrm{d}t$ with and without PID control.

Different colored $P_{\text{total}}(t)$ trajectories in Fig. \ref{Prealtime} correspond to different combinations of the PID gain tuple, shown in the top legends. In both sets (a.1)--(a.4) and (b.1)--(b.4), the real time aggregate consumption $P_{\text{total}}(t)$ trajectories, in general, show more oscillation than the optimal $P_{\text{total}}^{\text{ref}}(t)$, due to different time scales for operational planning (1 minute) and real time tracking (1 second); the gain tuple $(k_{p},k_{i},k_{d}) = (10^{-4}, 10^{-6}, 10^{-4})$ (shown in maroon), in general, performs better than others. Fig. \ref{Prealtime} (a.1)--(a.4) show almost perfect real time tracking for all four $\Delta$ distributions, since the movement of the comfort boundaries are unrestricted. In the presence of (\ref{TimeVaryingLowerBoundary}) and (\ref{TimeVaryingUpperBoundary}), as shown in (b.1)--(b.4), the tracking accuracy is limited compared to (a.1)--(a.4), and differs for different contractual comfort ($\Delta$) distributions.

\begin{figure}[t]
  \centering
    \includegraphics[width=0.48\textwidth]{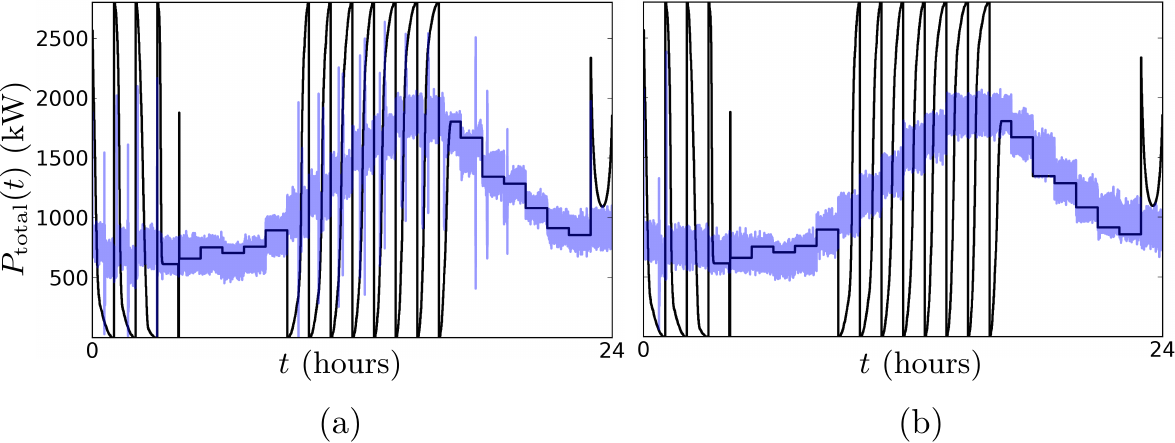}
    \caption{Tracking performance of the PID setpoint velocity controller with $(k_{p},k_{i},k_{d}) = (10^{-4}, 10^{-6}, 10^{-4})$, and black curves $P_{\text{total}}^{\text{ref}}(t)$ same as in Fig. \ref{Prealtime}(b.2) and \ref{NoisyTracking}, when individual consumption data at each time are (a) corrupted by $\pm 10\%$ i.i.d. uniform noise about the true values, and (b) both noise corrupted as in (a) as well as incomplete data reported due to communication by (each time randomly selected) 400 out of $N=500$ homes. The blue curves are the controlled real-time aggregate consumption. There is no significant difference in the tracking performance between those shown above and the same in Fig. \ref{Prealtime}(b.2) and \ref{NoisyTracking}.}
    \label{RobustTracking}
   \vspace*{-0.2in}
\end{figure}

In subfigure (b.1), all ACs have identical $\Delta \equiv 1^{\circ}$C, and hence identical $w_{\text{eff}}(t)$ for all $t$. Notice that about 16 hours onward, $P_{\text{total}}^{\text{ref}}(t)$ is decreasing, and to follow it in real time, the LSE moves all ACs' setpoints up at identical rates, resulting in all hitting their respective upper boundaries $U_{i0}$ simultaneously at around 20 hours. Between 20--23 hours, $w_{\text{eff}}(t)=0$ at $U_{i0}$, and $P_{\text{total}}(t)$ cannot track further decrease in $P_{\text{total}}^{\text{ref}}(t)$, as seen in subfigure (b.1). In subfigures (b.2)--(b.4), unlike (b.1), the setpoint boundaries for different $\Delta_{i}$ are moved at different rates. In (b.3) (resp. (b.4)), more ACs have large (resp. small) $\Delta_{i}$, thus allowing the LSE more (resp. less) flexibility in real time tracking, and result in less (resp. more) overshoots in $P_{\text{total}}(t)$. Comparing Fig. \ref{Prealtime}(b.2) with Fig. \ref{NoisyTracking}, it is seen that the privacy for individual consumptions can be preserved via local noise injection described in Section \ref{SensingTotalPowerSubsubection}, without affecting the tracking performance. Since the setpoint velocity control is reactive to error in the aggregate consumption, hence the tracking performance is robust against outliers, noise, and/or incompleteness in individual consumption data. In other words, the robustness observed in Fig. \ref{NoisyTracking} is typical in the sense that individual noise statistics or incomplete data does not significantly affect the aggregate tracking performance, as evidenced in Fig. \ref{RobustTracking}.


\begin{figure*}[t]
  \centering
    \includegraphics[width=0.9\textwidth]{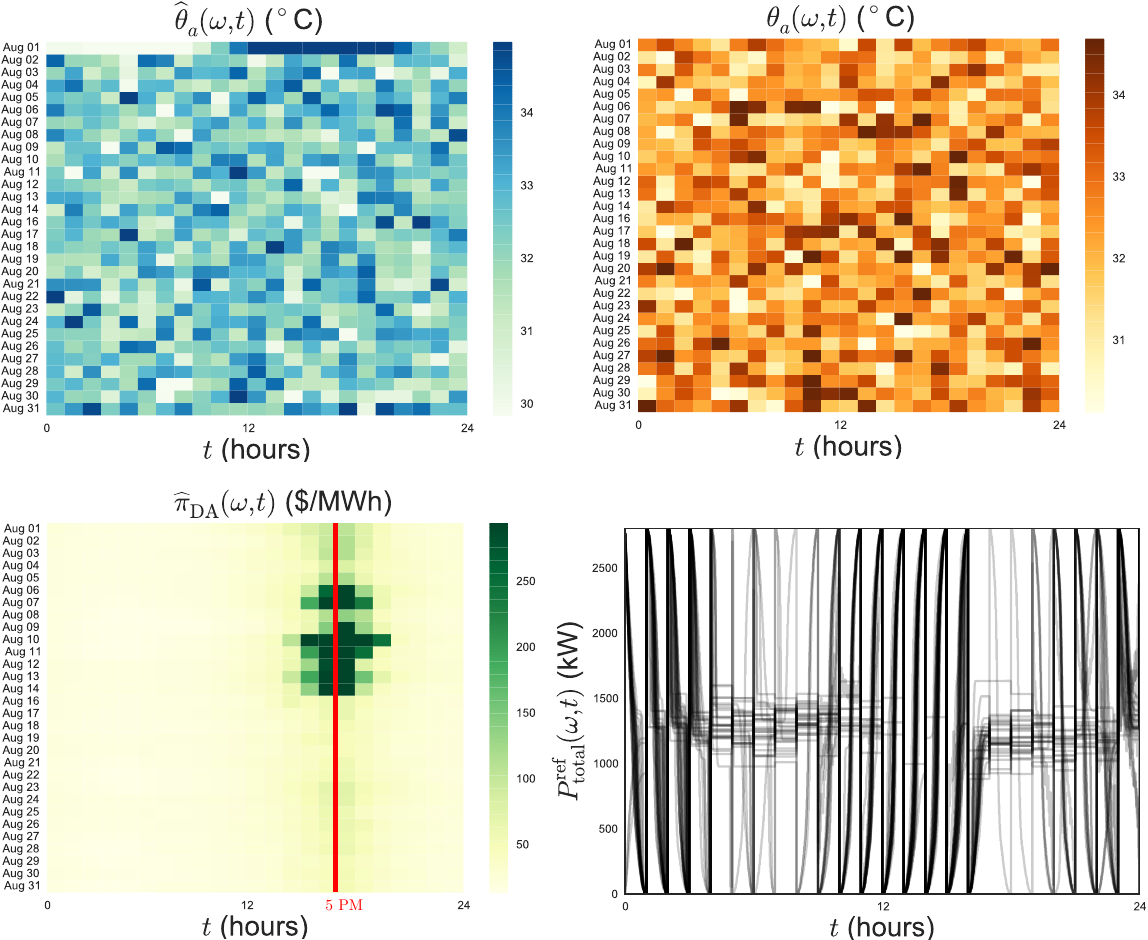}
    \caption{We use historical data for 30 days in August 2015, to generate 30 scenarios of $\widehat{\theta}_{a}(\omega,t)$ (top left), $\theta_{a}(\omega,t)$ (top right), $\widehat{\pi}_{\text{DA}}(\omega,t)$ (bottom left), for analyzing the performance of the LSE's strategy. The corresponding $P_{\text{total}}^{\text{ref}}(\omega,t)$ sample paths are shown in the bottom right.}
    \label{Scenario30data}
\end{figure*}

\begin{figure}[t]
  \centering
    \includegraphics[width=0.49\textwidth]{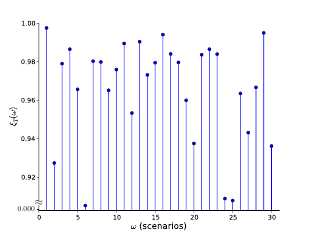}
    \caption{The normalized local time random variable $\xi_{T}(\omega)$ (see (\ref{NormalizedLocalTime})) is shown as a function of the 30 scenarios described in Fig. \ref{Scenario30data}.}
    \label{StatisticalPerformanceLocalTime}
\end{figure}

To demonstrate the statistical performance of the LSE's strategy, as discussed in Section \ref{PerformanceSubsection}, we take the historical data for forecasted day-ahead price and ambient temperature for 30 days in August 2015, and use these as the scenario set $\Omega$. One specific scenario $\omega\in\Omega$, refers to a trajectory tuple $(\widehat{\pi}_{\text{DA}}(\omega,t), \widehat{\theta}_{a}(\omega,t), \theta_{a}(\omega,t))$ that yields $P_{\text{total}}^{\text{ref}}(\omega,t)$ and $\xi_{T}\left(\omega\right)$ (see (\ref{NormalizedLocalTime})). For $\omega\in\Omega$, Fig. \ref{Scenario30data} shows all scenarios $(\widehat{\pi}_{\text{DA}}(\omega,t), \widehat{\theta}_{a}(\omega,t), \theta_{a}(\omega,t))$, and the corresponding $P_{\text{total}}^{\text{ref}}(\omega,t)$ for $\Delta$ uniformly distributed between $[0.1,1.1]^{\circ}$C with respective initial conditions and parameters mentioned in Section \ref{ParameterSubsubsection}, and $\overline{\tau}=0.45$ feasible for all $\omega\in\Omega$. Fig. \ref{StatisticalPerformanceLocalTime} shows that for most scenarios, the normalized local time $\xi_{T}\left(\omega\right)$ is close to unity, meaning that under the proposed strategy for the LSE, each AC would have zero effective widths for more than 90\% of times in a day, as expected in hot days in Houston.

\begin{figure}[t]
  \centering
    \includegraphics[width=0.49\textwidth]{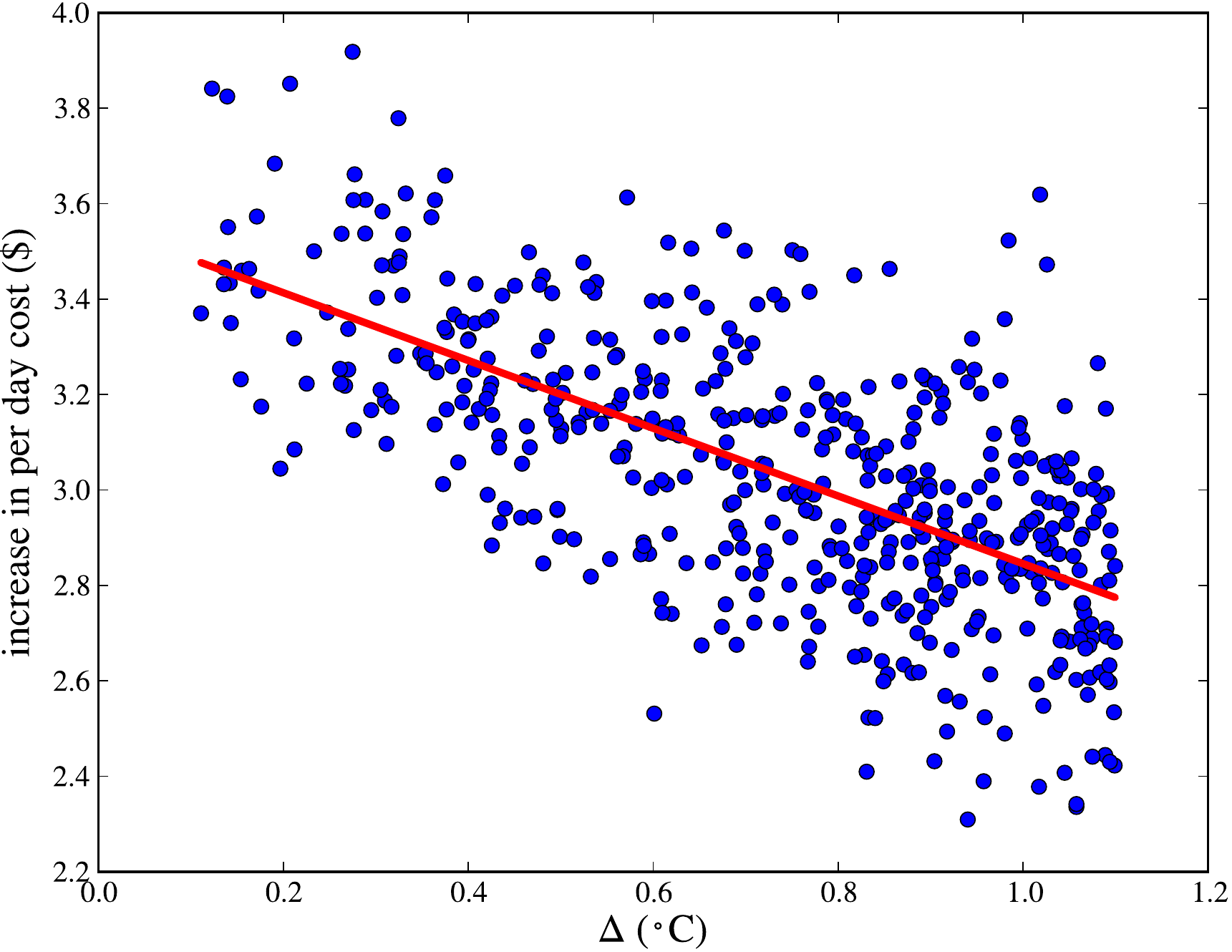}
    \caption{The scatterplot with blue circles denotes the increase in per day energy procurement cost, if an AC with comfort tolerance $\Delta_{i}$, is removed from the existing population of 500 ACs, as described in Section \ref{ContractSectionLabel}. The red line is the least squares linear fit that the LSE can use for pricing the contract of a new home.}
    \label{IndividualContract}
\end{figure}

In Fig. \ref{IndividualContract}, we plot results for the sensitivity analysis described in Section \ref{ContractSectionLabel}, with the data shown in Fig. \ref{PriceAmbient}, for $\Delta$ uniformly distributed in $[0.1, 1.1]^{\circ}$C. The red line in Fig. \ref{IndividualContract} is the least squares linear fit for the scattered data, and can be used to price the contract with a new home for its chosen comfort tolerance $\Delta$. As expected intuitively, a smaller $\Delta$ leads to higher per day service charge, and vice versa.

\subsection{Computational Complexity}
\label{ComputationalComplexitySectionLabel}
The computational complexity of the proposed method is twofold: that of the \emph{optimal planning}, and that of the \emph{setpoint control}. The first being an \emph{offline} computation, scales as a function of both the number of homes $N$, and the discretization-length $\mu := \ceil[\big]{\frac{T}{\Delta t}}$ of the time horizon $[0,T]$, where $\Delta t$ is the discretization step-size for the ODE (\ref{DynamicsConstraint}). The setpoint control being an \emph{online} computation, scales as a function of $N$.

\begin{figure}[h]
  \centering
    \includegraphics[width=0.45\textwidth]{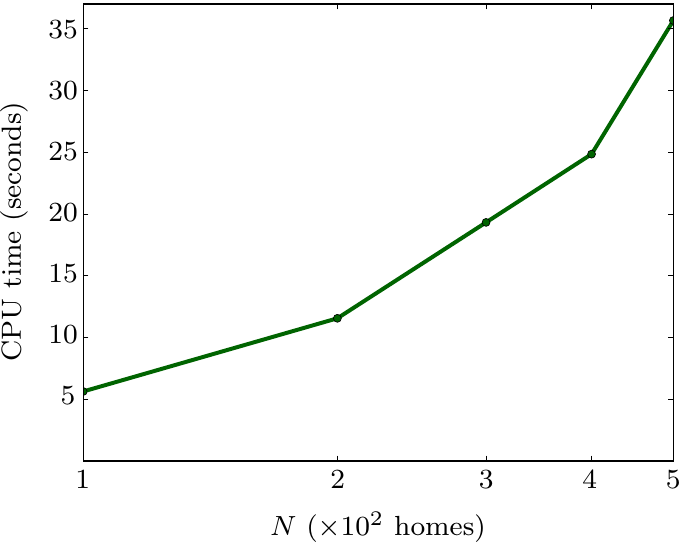}
    \caption{CPU time (in seconds) needed to execute MATLAB \texttt{linprog} for solving the planning problems for $N = 100, 200, 300, 400$ and $500$ homes, corresponding to case (iv) in Section \ref{ActualDataSubsubsection}. The CPU time data for other cases are almost identical to the one shown, and omitted for brevity. The horizontal axis is in logarithmic scale while the vertical axis is in linear scale.}
    \label{NvsCPUtimeLinprog}
\end{figure}

As described in Section \ref{MILPvsLPsubsubsection}, Euler discretization for ODE (\ref{DynamicsConstraint}) leads to an LP in decision vector $\{\theta_{1}(1), u_{1}(1), \theta_{2}(1),$ $u_{2}(1), \hdots, \theta_{N}(1), u_{N}(1), \hdots, \theta_{N}(\mu), u_{N}(\mu)\}$ of length $2N\mu$ with $N\mu+1$ equality constraints and $2N\mu$ inequality constraints. Here, $\theta_{i}\left(k\right)$ and $u_{i}(k)$ respectively denote the indoor temperature and optimal control for the $i$\textsuperscript{th} home at time $t=k\Delta t$, where $i=1,\hdots,N$, and $k=1,\hdots,\mu$. We solved this LP via MATLAB \texttt{linprog} using Mehrotra's implementation of primal-dual interior point solver \cite{Mehrotra1992}, that has worst-case complexity polynomial in $N\mu$. For our particular instance of linear program and for $\Delta t = 1$ minute (i.e., $\mu = 1440$), the runtime complexity scaling with $N$ is shown in Fig. \ref{NvsCPUtimeLinprog}. It is easy to verify that executing equations (\ref{SetpointVelocityControl}), (\ref{TimeVaryingLowerBoundary}) and (\ref{TimeVaryingUpperBoundary}) for real-time setpoint control, requires $\mathcal{O}(N)$ runtime complexity.

The LPs for computing optimal reference power consumption trajectories, were solved in MATLAB R2015a running on Dell PC with 3.4 GHz Intel i7-2600 processor and 16 GB memory. The same platform was used for solving the MILP mentioned in Section \ref{MILPvsLPsubsubsection} using Gurobi. All other simulations including setpoint velocity control were performed in MATLAB R2013a running on a Macbook Pro with 2.6 GHz Intel Core i5 processor and 8 GB memory.


\section{Conclusion}
We have presented an architecture and control algorithms for an aggregator or load serving entity managing a population of thermal inertial loads. The proposed hierarchical architecture solves the twin problems of minimum cost energy procurement from the day-ahead market, and real-time setpoint control to track the reference aggregate power trajectory that corresponds to the optimal energy procurement. The control algorithms respect privacy and contractual comfort at the individual consumer level, but can be implemented at the aggregate level. We demonstrate these algorithms using the day-ahead price data from ERCOT, and using the forecasted and real time ambient temperature data from a weather station in Houston, Texas. Future work will investigate the potential of such contracts as innovative business models for load serving entities.


\appendices

\section{Differentially Private Estimate of Total Power}
\label{AppDiffPrivacy}
At time $t$, let us consider two consumption vectors $\mathbf{c}(t), \mathbf{c}^{\prime}(t) \in \{0,P_{e}\}^{N}$. We call two consumption vectors \emph{neighboring} if they differ by at most one element. Then, for any neighboring consumption pair $\mathbf{c}(t), \mathbf{c}^{\prime}(t)$, and for sum query $\text{SUM} : \{0,P_{e}\}^{N} \mapsto \{0,P_{e},2P_{e},\hdots,NP_{e}\}$, we have $\sup_{\mathbf{c}(t),\mathbf{c}^{\prime}(t)} | \text{SUM}(\mathbf{c}(t)) - \text{SUM}(\mathbf{c}^{\prime}(t)) | \leq P_{e}$. Hence, $\epsilon$-differential privacy \cite{DworkRoth2014} at the LSE level is guaranteed if $\widetilde{P}_{\text{total}}(t) = \sum_{i=1}^{N}P_{i} + n$, where $n \sim \text{Lap}\left(\frac{P_{e}}{\epsilon}\right)$. 
%

To see that the strategy proposed in Section \ref{SensingTotalPowerSubsubection} achieves $\epsilon$-differential privacy, notice that $n_{i} \sim \text{Gamma}(\frac{1}{pN},\frac{\epsilon}{pP_{e}}) \Rightarrow \frac{1}{p}n_{i} \sim \text{Gamma}(\frac{1}{pN},\frac{\epsilon}{P_{e}})$, and since $\widehat{N} = pN$, we have $\widetilde{n} := \sum_{i=1}^{\widehat{N}}\frac{1}{p}n_{i} \sim \text{Exp}(\frac{\epsilon}{P_{e}})$. We thus arrive at
\begin{eqnarray}
\widetilde{P}_{\text{total}}(t) = \widehat{P}_{\text{total}}(t) - \nu = \frac{1}{p}\sum_{i=1}^{\widehat{N}}P_{i}(t) + (\widetilde{n} - \nu),
\end{eqnarray}
where the first term $\frac{1}{p}\sum_{i=1}^{\widehat{N}}P_{i}(t)$ provides an estimate for the true aggregate consumption $\sum_{i=1}^{N}P_{i}(t)$; the second term $n:= \widetilde{n} - \nu \sim \text{Lap}(\frac{P_{e}}{\epsilon})$ since $\widetilde{n}, \nu$ are i.i.d. $\text{Exp}(\frac{\epsilon}{P_{e}})$.



\section{Deriving equation (\ref{EffectiveDeadbandWidth})}
\label{SomeAlgebra}
For $a,b,x\in\mathbb{R}$, recall that
\begin{eqnarray}
(a + x) \wedge (b + x) &=&(a \wedge b) + x, \label{MinOfSum}\\
(a - x) \vee (b - x) &=& (a \vee b) - x, \label{MaxOfDifference}\\
-\left(0 \wedge a\right) &=& [-a]^{+}, \label{NegativeOfMinWithZero}
\end{eqnarray}
and hence we can rewrite (\ref{TimeVaryingLowerBoundary}) and (\ref{TimeVaryingUpperBoundary}) as
\begin{eqnarray}
L_{it} &\stackrel{(\ref{MaxOfDifference})}{=}& \left(s_{i0} + \Delta_{i}\right) \wedge \left[\left(s_{i0} \vee s_{i}(\omega,t)\right) - \Delta_{i}\right] \nonumber\\
&\stackrel{(\ref{MinOfSum})}{=}& \{s_{i0} \wedge \left[\left(s_{i0} \vee s_{i}(\omega,t)\right) - 2\Delta_{i}\right]\} + \Delta_{i},	
\label{LowerBndry}\\
U_{it} &\stackrel{(\ref{MinOfSum})}{=}& \left(s_{i0} - \Delta_{i}\right) \vee \left[\left(s_{i0} \wedge s_{i}(\omega,t)\right) + \Delta_{i}\right] \nonumber\\
&\stackrel{(\ref{MaxOfDifference})}{=}& \{s_{i0} \vee \left[\left(s_{i0} \wedge s_{i}(\omega,t)\right) + 2\Delta_{i}\right]\} - \Delta_{i}.	
\label{UpperBndry}
\end{eqnarray}
This results $w_{\text{eff}}(i,\omega,t) := U_{it} - L_{it} = T_{1} - T_{2} - 2\Delta_{i}$, where $T_{1} := \{s_{i0} \vee \left[\left(s_{i0} \wedge s_{i}(\omega,t)\right) + 2\Delta_{i}\right]\}$, and $T_{2} := \{s_{i0} \wedge \left[\left(s_{i0} \vee s_{i}(\omega,t)\right) - 2\Delta_{i}\right]\}$.

If $\int_{0}^{t}v(\omega,\varsigma)\mathrm{d}\varsigma > 0$, then $s_{i}(\omega,t) > s_{i0}, \Rightarrow T_{1} = s_{i0} \vee \left[s_{i0} + 2\Delta_{i}\right] = s_{i0} + 2\Delta_{i}$, and $T_{2} = s_{i0} \wedge \left[s_{i}(\omega,t) - 2\Delta_{i}\right] = s_{i0} \wedge \left[s_{i0} + \Delta_{i}\int_{0}^{t}v(\omega,\varsigma)\mathrm{d}\varsigma - 2\Delta_{i}\right] \stackrel{(\ref{MinOfSum})}{=} s_{i0} + \Delta_{i}\{0 \wedge [\int_{0}^{t}v(\omega,\varsigma)\mathrm{d}\varsigma - 2]\}$. This yields 
\begin{eqnarray}
w_{\text{eff}}(i,\omega,t) &=& -\Delta_{i}\left(0 \wedge \left[\int_{0}^{t}v(\omega,\varsigma)\mathrm{d}\varsigma - 2\right]\right) \nonumber\\
&\stackrel{(\ref{NegativeOfMinWithZero})}{=}& \Delta_{i}\left[2 - \int_{0}^{t}v(\omega,\varsigma)\mathrm{d}\varsigma\right]^{+}.
\label{PositiveIntegralEffWidth}
\end{eqnarray}

On the other hand, if $\int_{0}^{t}v(\omega,\varsigma)\mathrm{d}\varsigma \leq 0$, then $s_{i}(\omega,t) \leq s_{i0}, \Rightarrow T_{1} = s_{i0} \vee (s_{i}(\omega,t) + 2\Delta_{i}) = s_{i0} \vee (s_{i0} + \Delta_{i}\int_{0}^{t}v(\omega,\varsigma)\mathrm{d}\varsigma + 2\Delta_{i})$, and $T_{2} = s_{i0} \wedge (s_{i0} - 2\Delta_{i}) = s_{i0} - 2\Delta_{i}$, resulting 
\begin{eqnarray}
w_{\text{eff}}(i,\omega,t) 
&=& [s_{i0} \vee (s_{i0} + \Delta_{i}\int_{0}^{t}v(\omega,\varsigma)\mathrm{d}\varsigma + 2\Delta_{i})] - s_{i0} \nonumber\\
&\stackrel{(\ref{MaxOfDifference})}{=}& \Delta_{i}\left[2 + \int_{0}^{t}v(\omega,\varsigma)\mathrm{d}\varsigma\right]^{+}.
\label{NegativeIntegralEffWidth}
\end{eqnarray}
Combining (\ref{PositiveIntegralEffWidth}) and (\ref{NegativeIntegralEffWidth}), we arrive at (\ref{EffectiveDeadbandWidth}).



%

%



\vspace*{-0.5in}

\begin{IEEEbiography}[{\includegraphics[width=1in,height=1.25in,clip,keepaspectratio]{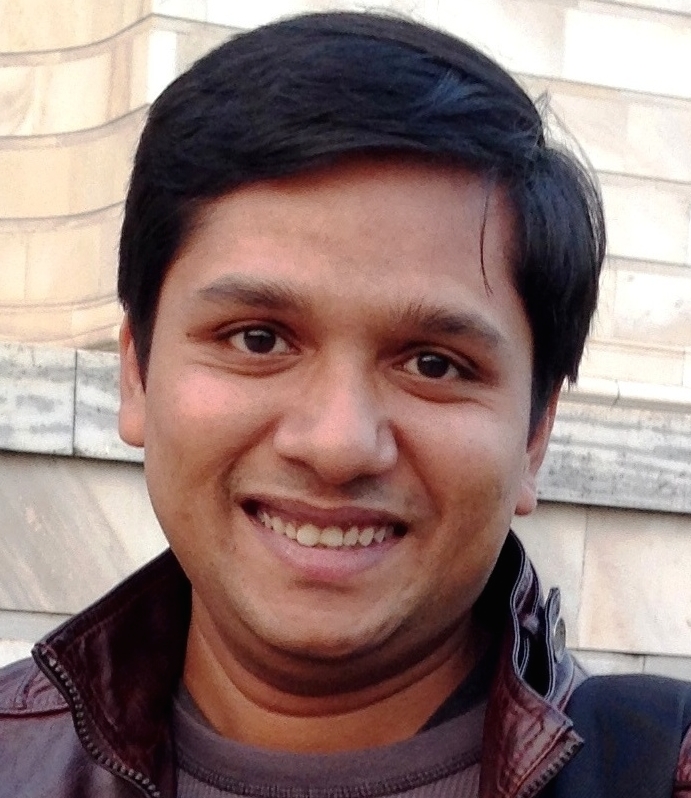}}]{Abhishek~Halder} 
(S'10-M'14) received the B. Tech. and M. Tech. degrees in aerospace engineering from IIT Kharagpur in 2008, and the Ph.D. degree in aerospace engineering from Texas A\&M, College Station, TX, USA, in 2014. He is a Postdoctoral Research Associate at the Department of Electrical and Computer Engineering, Texas A\&M University. His research interest is in systems, control and optimization with focus on large-scale cyberphysical systems such as the smart grid and aerial swarm robotics. He received the 2014 Outstanding Doctoral Student Award from Texas A\&M and the 2008 Best Thesis Award from IIT, both in Aerospace Engineering.
\end{IEEEbiography}


\begin{IEEEbiography}[{\includegraphics[width=1in,height=1.25in,clip,keepaspectratio]{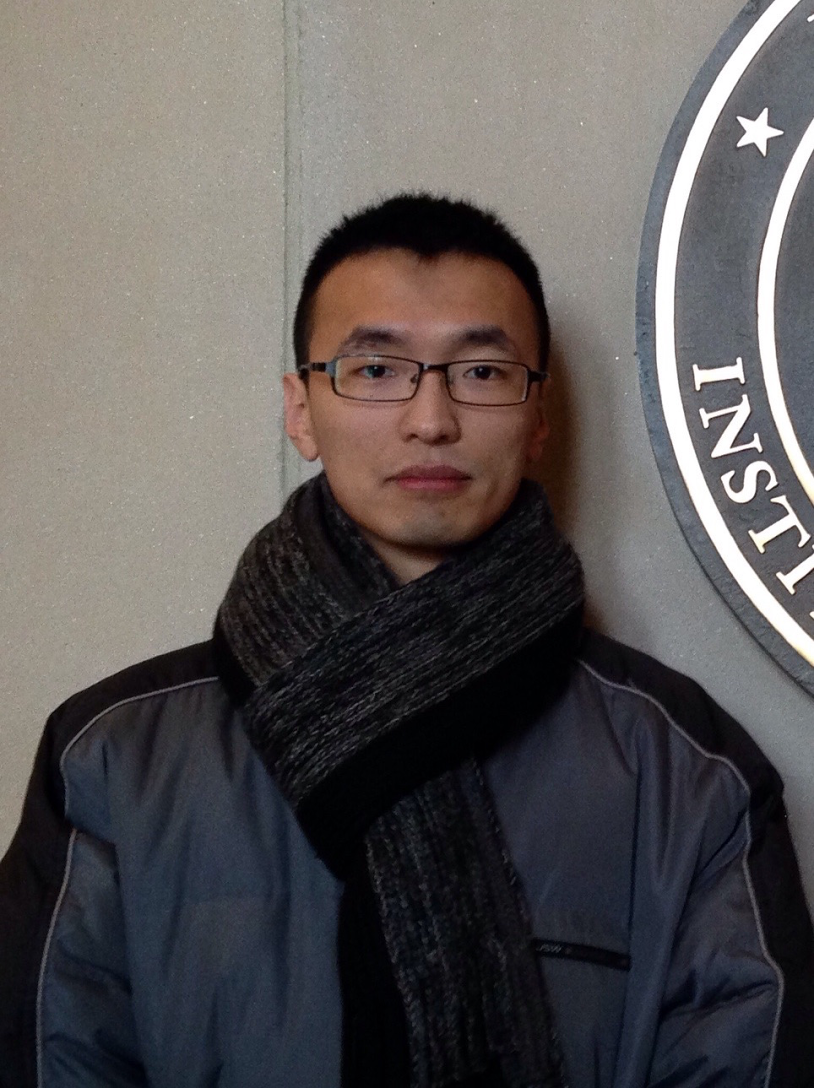}}]{Xinbo~Geng} 
(S'14) received the B.E. degree in electrical engineering in 2013 from Tsinghua University, Beijing, China, and the Master of Science degree from the Department of Electrical and Computer Engineering at Texas A\&M University in 2015, where he is currently working toward the Ph.D. degree. His research interests include demand response, electricity markets and data analytics in power system.
\end{IEEEbiography}


\begin{IEEEbiography}[{\includegraphics[width=1in,height=1.25in,clip,keepaspectratio]{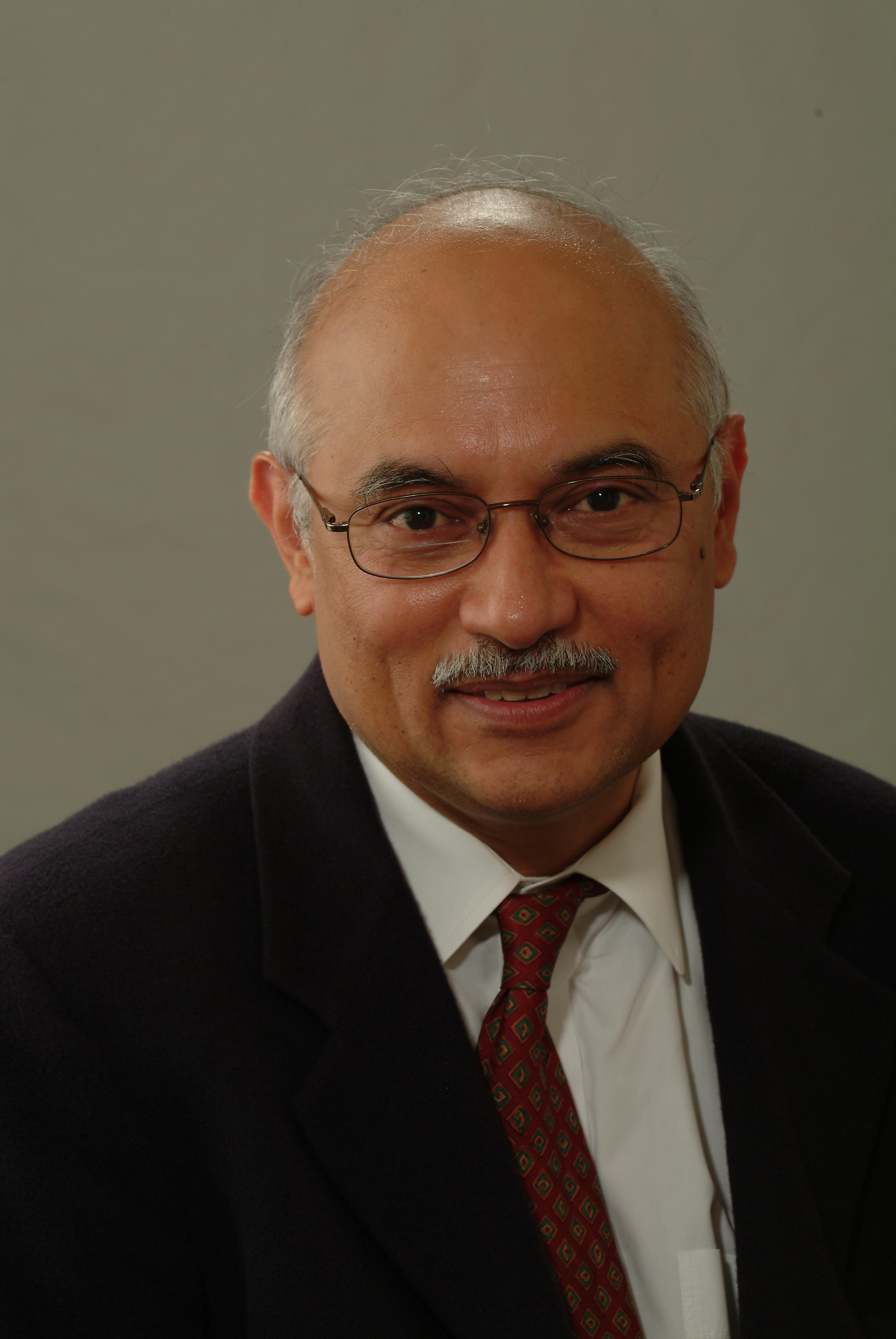}}]{P.~R.~Kumar} 
(F'88) received the B. Tech. degree from IIT Madras, Chennai, India, in 1973, and the D.Sc. degree from Washington University, St. Louis, MO, USA, in 1977. He was a faculty member at UMBC (1977–1984) and University of Illinois, Urbana-Champaign (1985–2011). He is currently at Texas A\&M University. His current research is focused on stochastic systems, energy systems, wireless networks, security, automated transportation, and cyberphysical systems. 

He is a member of the US National Academy of Engineering and The World Academy of Sciences. He was awarded a Doctor Honoris Causa by ETH, Zurich. He has received the IEEE Field Award for Control Systems, the Donald~P.~Eckman Award of the AACC,  Fred~W.~Ellersick Prize of the IEEE Communications Society, the Outstanding Contribution Award of ACM SIGMOBILE, the IEEE Infocom Achievement Award, and the SIGMOBILE Test-of-Time Paper Award. He is a Fellow of IEEE and an ACM Fellow. He was Leader of the Guest Chair Professor Group on Wireless Communication and Networking at Tsinghua University, is a D. J. Gandhi Distinguished Visiting Professor at IIT Bombay, and an Honorary Professor at IIT Hyderabad. He was awarded the Distinguished Alumnus Award from IIT Madras, the Alumni Achievement Award from Washington Univ., and the Daniel Drucker Eminent Faculty Award from the College of Engineering at the Univ.~of Illinois.
\end{IEEEbiography}

\vfill


\begin{IEEEbiography}[{\includegraphics[width=1in,height=1.25in,clip,keepaspectratio]{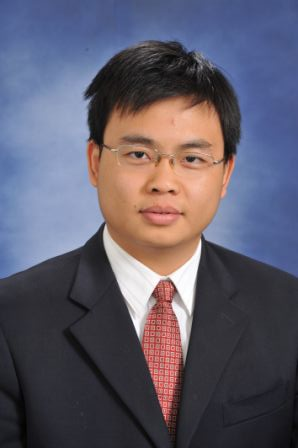}}]{Le~Xie} 
(M'10-SM'16) received the B.E. degree in electrical engineering from Tsinghua University, Beijing, China, in 2004, the S.M. degree in engineering sciences from Harvard University, Cambridge, MA, USA, in June 2005, and the Ph.D. degree from the Electric Energy Systems Group (EESG), Department of Electrical and Computer Engineering, Carnegie Mellon University, Pittsburgh, PA, USA, in 2009. He is an Associate Professor in the Department of Electrical and Computer Engineering at Texas A\&M University, where he is affiliated with the Electric Power and Power Electronics Group. His research interest includes modeling and control of large scale complex systems, smart grid applications in support of renewable energy integration, and electricity markets.
\end{IEEEbiography}





\vfill

\enlargethispage{-5in}

\end{document}